\documentclass[12pt, preprint]{aastex}
\shorttitle{Dynamical evolution of molecular clouds}
\begin{document}
\title{An Investigation on the Morphological Evolution of Bright-Rimmed Clouds(BRCs)}

\author{Jingqi Miao\altaffilmark{1}, Glenn, J. White\altaffilmark{2,3},
 M. A. Thompson\altaffilmark{4}, Richard P. Nelson\altaffilmark{5}}

\altaffiltext{1}{Centre for Astrophysics \& Planetary Science, School of Physical Sciences, University of
Kent, Canterbury, Kent CT2 7NR, UK, 
J.Miao@kent.ac.uk}
\altaffiltext{2}{Centre for Earth, Planetary, Space \& Astronomical Research, The Open University,
Walton Hall, Milton Keynes, MK7 6AA }

\altaffiltext{3}{Space Physics Division, Space Science \& Technology Division,
CCLRC Rutherford Appleton Laboratory, Chilton, Didcot, Oxfordshire, OX11 0QX, UK}

\altaffiltext{4}{School of Physics Astronomy \& Maths, University of Hertfordshire, College Lane, Hatfield,
AL10 9AB, UK}

\altaffiltext{5}{School of Mathematical Sciences, Queen Mary College, University of London, Mile End Road,
London E1 4NS, UK}

\begin{abstract}
A new Radiative Driven Implosion (RDI) model based on  Smoothed Particle
Hydrodynamics (SPH) technique is developed and applied to investigate the morphological
evolutions of molecular clouds under the effect
of ionising radiation. This model self-consistently
includes the self-gravity of the cloud in the hydrodynamical evolution, the UV radiation 
component in the radiation transferring equations, the relevant heating and cooling mechanisms
in the energy evolution and a comprehensive chemical network. The simulation results 
reveal that under the effect of ionising radiation,
a molecular cloud may evolve through different evolutionary sequences. 
Dependent on its initial gravitational state, the evolution of a molecular cloud 
does not necessarily follow a complete morphological evolution sequence 
from type A$\rightarrow$B$\rightarrow$C, as
described by previous RDI models. When confronted with observations, the simulation results
provide satisfactory physical explanations for a series of puzzles derived 
from Bright-Rimmed Clouds(BRCs) observations.
The consistency of the modelling results with observations shows that the self-gravity of a molecular cloud
should not be neglected in any investigation on the dynamical evolution of molecular clouds when they are
exposed to ionising radiation.       
\end{abstract}


\section{Introduction}
Bright-rimmed clouds (BRCs) found in and around HII regions are sites of ongoing star formation due to 
compression by ionisation/shock fronts, and
provide an excellent laboratory to study the influence of UV radiation from nearby massive stars on the
evolutionary process of molecular clouds. Numerous molecular line, millimetre/sub-millimetre continuum 
and mid-IR surveys have revealed very detailed structures and physical properties of BRCs in various 
astrophysical environments \citep{Sugitania, Sugitanib, Thompsona, Thompsonb, Thompsonc, Urquhart}.

One of the most intriguing characteristics of the observed BRCs is the diversity of their
morphologies. The observed BRCs appear in different morphologies even  when they are in
similar stellar environments. Sugitani et al (1991, 1994) classified 89 BRCs from a whole sky survey
into three types depending on the curvature of the rim of BRCs. The three types are categorised  with 
type A cloud with a rim displaying moderate curvature, type B cloud with a rim of a high degree of curvature 
which sometimes is also described as an elephant trunk morphology, 
and finally type C cloud with a tightly curved rim 
and a tail, which is also called cometary globule.  
 
\citet{lefloch} presented a 2-dimensional numerical simulation of the effect of UV radiation on the 
dynamical evolution of a 
molecular cloud based on the Radiation Driven Implosion (RDI) model.     
Although their hydrodynamical modelling successfully created 
a complete sequence of the morphological evolution of a molecular cloud from type A to B and then to C,  
it seems difficult to provide physical explanations for the following questions derived 
from BRC observations:
 
\noindent 1) Why are there so many more type A BRCs? 
The first whole sky survey by Sugitani et al (1991,1994) 
revealed that 61\% of 
the observed 89 BRCs have type A morphology. If all of the clouds followed the same evolutionary sequence 
revealed by Lofloch's modelling, from type A to type C morphology, there should be a balanced number of
BRCs with different morphologies to be observed.   

\noindent 2) Why do some of observed BRCs evolve to a quasi-stable type A rimmed morphology and 
 show signs of star formation at the heads of their structures?   
If all of the clouds should evolve to type C morphology and finally were completely 
photoevaporated as described by Lelfloch's modelling, 
no star formation should be triggered in type A BRCs. 
The recent radio continuum
and molecular line observations on BRCs  in the catalogue of SFO 
\citep{Sugitania, Sugitanib} presented by \citet{Urquhart} and 
many other observations in molecular lines and submilimeter lines \citep{Lee, Karr, Thompsonb} 
provided strong evidences of ongoing star formations in the condensed head of type A rimmed clouds where
a pressure equilibrium at the boundary between the clouds and its environment
 has been reached. Therefore it seems that for some of the molecular clouds, 
further evolution to type B or C rimmed morphologies is not necessary, i.e., not every BRC's
morphological evolution follows the sequence from type A to B then to C rimmed morphology, it    
may terminate their morphological evolution at any one of the three morphologies. 

\noindent 3) What causes the spatial morphological distribution of BRCs in some of molecular cloud clusters? 
The remnant molecular clouds in the Ori OB 1 associations \citep{ogura} presented a spatial 
sequence from type A$\rightarrow$B$\rightarrow$C rimmed BRCs with their distance from the ionising star. 
It was  once suggested that some molecular clouds which were farther away may directly evolve 
to type C rimmed morphology \citep{ogura}, but there is no physical foundation for this speculation.  

For the molecular clumps very close to an ionising star, the ionised gas cannot expand freely from
the surface of the clump, because their expansion is constrained by a material flow accompanying with 
the ionisation radiation flux. Lefloch's simulation showed that a zero radial velocity boundary condition,  
(the extreme case of flow effect) favours the type A BRC formation \citep{lefloch}. 
This result may partly explain the predominance of type A BRCs at the border of the HII regions.
However question remains for the type A BRC structures found at the outer location of the HII regions
or type B/C BRCs found at the border of HII regions \citep{Sugitania,Sugitanib} and there are must be
some other physical mechanisms which  play dominant roles in the morphological evolution of molecular clouds.                
   
Therefore searching for satisfactory solutions to the above puzzles requires a more 
comprehensive model which                    
should include as many physical processes as possible so that  
reliable physical explanations on what observations revealed can be derived, hence to greatly improve 
our understanding on how the intensive UV radiation from nearby stars affects the 
dynamical evolution of the surrounding molecular clouds.

The fact that IRAS sources and bipolar outflows in cloud globules are very often found at the edges of BRCs
strongly suggests that the self-gravity of a molecular cloud can play an active role even in the early stages 
of evolution in molecular clouds 
\citep{lefloch}. Nevertheless the self-gravity of a molecular cloud was neglected in Lefloch's model, because
 the initiative of the model  was to investigate  photo-evaporation effect of molecular clouds, 
for which self-gravity does not play a dominant role. 
However, as pointed out by \citet{lefloch} at 
the end of the discussions of their
modelling results, inclusion of the self-gravitation of a molecular cloud is necessary  in order to build
a appropriate model for investigation on the dynamical evolution of globule clouds under the effect of UV
radiation. 

Recently, \citet{Kessela, Kesselb}
have developed the first 3-D SPH RDI model for investigation of the effect of UV radiation on the dynamical 
evolution of molecular  
clouds. Their model includes both self-gravity and the hydrogen ionisation 
of the molecular cloud in the dynamical evolution equations and revealed further features
of molecular cloud evolution under the effect of ionising radiation. However a self-consistent treatment 
for the energy evolution was not included  which
sets a barrier for a rigorous study on the evolution of
the physical properties of a cloud, so that a 
direct confrontation with observations is difficult.  

Therefore it is our intention  to build a
comprehensive RDI model
in order to obtain a consistent description about the effect of the UV radiation on the 
dynamical evolution  of a molecular cloud and to reveal the physical origin for the 
observed characteristics of BRCs' morphology formation.    

In the following sections we first give an outline for our 3-D RDI model, then briefly present
an analytical solution which uses some observable physical properties
of a condensed globule and the UV radiation flux to define a maximum mass a 
stable condensed globule could have to against the gravitational collapse. Next the dynamical evolution
of the cloud will be detailed and the corresponding kinematics will be analysed based on numerical 
simulation results. Finally we will apply the derived
knowledge to confront with BRCs' observations and the conclusions are drawn at the end of the paper. 

\section{The model}
In this section, we present a brief description of the basic equations and various physical processes 
included in our model. The hydrodynamical equations are solved with Smoothed Particle Hydrodynamical(SPH) technique. 
For the  detailed description of the SPH theory and the corresponding technique,
readers are directed to \citet{nelson}.
\subsection{The basic dynamical equations}
The continuity, momentum, and energy equations for a compressible fluid can be written as:
\begin{equation}
\frac{d \rho}{d t} + \rho \nabla \cdot \bf{v}  =  0
\end{equation}
\begin{equation}
\frac{d \bf{ v}}{d t}  =  - \frac{1}{\rho} \nabla P - \nabla \Phi + \bf{S}_{visc}
\end{equation}
\begin{equation}
\frac{d \cal{U}}{d t} + \frac{P}{\rho} \nabla \cdot \bf{v}   =  \frac{\gamma - \Lambda}{\rho}
\end{equation}
and the chemical rate equations take the general form:
\begin{equation}
\frac{d X_i}{d t}=n K_i
\label{Xi}
\end{equation}
where $ d / d t = \partial / dt + \bf{v} \cdot \nabla$ denotes the convective derivative,
$\rho$ is the density; $\bf{v}$ is the velocity; $P$ is the pressure; $\bf{S_{visc}}$ represents the viscous
forces; $\cal {U}$ is the internal energy per unit mass; and $\Phi$ is the gravitational potential.
${\gamma}$ and ${\Lambda}$ represent the non adiabatic heating and cooling functions respectively, which we will
 deal with later;
The fractional abundance of  main chemical species $X_i$ in
Equation (\ref{Xi}) are  for CO, CI, CII,
HCO$^+$, OI, He$^+$, H$_3^+$,
OH$_{x}$, CH$_x$, M$^+$ and electrons; $K_i$ is the associated chemical
reaction rate and $n$ is the total  number density. 

\subsection{Ionising radiation transfer}
The molecular cloud in our model is under the effects of both UV  and FUV radiation fields. 
The UV radiation is from a nearby star and  interacts with the molecular cloud through  the 
top layer of the surface facing the star (front surface). 
The FUV radiation onto the front surface is from the nearby star and that onto the rear surface 
is from the background star light \citep{nelson}. Then the FUV radiation 
field around a molecular cloud can be approximated as a spherical radiation field \citep{gorti}.
The the effect of the FUV field on the molecular cloud is mainly through the photoelectric 
emission of electrons from grains 
and is treated in the same way as that in  \cite{nelson}. In the following part of this subsection
we only need discuss how to deal with the ionising radiation (UV) field from the nearby star.     
    
Although Helium ionisation was included in the chemical network, we could safely neglect it when
dealing with the ionisation radiation transfer, ionisation heating and cooling for simplicity, because of
the much lower abundance of the Helium compared to hydrogen atoms \citep{Dyson}.
The implementation of ionising radiation from nearby stars into the above
SPH code is based on solving the following ionisation rate and radiative transfer equations,
\begin{eqnarray}
\frac{d n_e}{d t}  & = & \mathcal{I} - \mathcal{R} \\
\label{ne}
\frac{d J}{d z} &  = & - \sigma n (1 - x) J
\label{J0}
\end{eqnarray}
where ${\mathcal{I}} ={\sigma} n (1 - x) J$ is the
ionisation rate with $\sigma$ being the effective ionisation cross section; 
${\mathcal{R}} = n_e^2 \alpha_B = x^2 n^2 \alpha_B$ is
the recombination rate with $\alpha_B$ being the
effective recombination coefficient under the assumption of the 'on the spot' approximation
\citep{Dyson}. In its original definition, the recombination coefficient
\[\alpha = \sum_i \alpha_i \] includes all of the individual recombination coefficients $\alpha_i$
to  the hydrogen atomic energy level $i$. In the 'on the spot' assumption, recombinations into the ground
level ($ i=1$) do not lead to any net effect on the change in ionisation rate,
since the photons released from this recombination process  are able to re-ionise other hydrogen atoms
on the spot. Therefore $\alpha_1$ can be neglected and the resulting net recombination coefficient can
be written as \[ \alpha_B = \sum \limits_{i=2}^{\infty} \alpha_i \]
For a planner infall of ionising photons from a distant source on the boarder  with a flux $J_0$
as the Lyman continuum photons per unit time and square area, the solution of Eq.({\ref{J0}}) is
\begin{displaymath}
J(z)=J_0 \, exp[- \tau(z)]
\end{displaymath}
where $\tau(z)$ is the optical depth for ionising photons along the line of
sight parallel to the infall direction of the photons, and $z$ is the distance from
the border of the integration volume along the line of sight. Neglecting the absorption
effect of dust on ionising radiation, $\tau(z)$ can be expressed as \citep{Kessela}
\begin{equation}
\tau(z)=\int^{\, z}_R n_H (z') \, \, \bar{\sigma} \,\, dz'
\end{equation}
with $n_H = n ( 1 - x )$ and $\bar{\sigma}$ being the mean of $\sigma_v$ over frequency, weighted
by the spectrum of the source \citep{Kessela}. 
                                                                                
Eq.(\ref{ne}) can be written in terms of the ionisation ratio $x$, i.e.,
\begin{equation}
\frac{d x}{dt} \, \,= \, \frac{1}{n}\, \frac{dn_e}{dt}= \frac{\mathcal{I}}{n} -  x^2 n \alpha_B
\label{J1}
\end{equation}
Solving Eq.({\ref{J1}) for the ionising ratio $x$ includes three steps: i.e., calculate the optical
path $\tau(z)$; calculate the ionising rate $I$ and then finally integrate Eq.(\ref{J1}) for $x$.
We employ the method introduced by \citet{Kessela} in the above three-step work. We just give
a very short description about it here. The interested reader can find more detailed description in
\citet{Kessela}.
                                                                                                                           
For any a simulated particle $i$ with its number density $n$ and its nearest neighbour list being given by the
dynamical simulation at each time step, calculating $\tau(z)$ is carried through
a) finding the evaluation points on the path toward source by smallest angel criteria and then
b)summing up all the optical path elements on the path
\[\tau(z_i)= \Sigma_{k=0}^{m} \, \, \bar{\sigma} \, \,  n_{H,k} \, \, \Delta z_k \  \]
where $m$ is the total number of the evaluation points from the boundary of the volume to the
position located one effective radius ($a_i = 0.85 (M_i / \rho_i )^{1/3} $) before the
point $z_i$ which is derived from a);
 $n_{H,k}$ is the number density of the hydrogen atoms at the
evaluation point $k$, $M_i$ and $\rho_i$  are the mass and mass density of the particle $i$ respectively.

The ionisation rate for the particle $i$ is then calculated by
\[I_i = \frac{J_0}{2 a_i n_i} exp [- \tau(z_i)] \,\, [1 - exp (- 2 a_i \, n_{H,i})] \]
Finally,  the first-order discretisation of Eq. (\ref{J1}) over a time interval $\Delta t$ for particle $i$
 is given by
\begin{equation}
x_i^{j+1} = x_i^j + \Delta t \, \, [ \frac{\mathcal{I}_i^{j+1}}{n_i^{j+1}} - n_i^{j+1} \, (x_i^2)^{j+1} \alpha_B ]
\end{equation}
where the indexes $j$ and $j+1$ denote the values at the beginning and the end of the actual
time-step $\delta t$ respectively. All of the values on the right-hand side are known from advancing
the particles by the SPH formalism, except $\mathcal{I}_i^{j+1}$, which can be approximated by
\begin{equation}
\mathcal{I}_i^{j+1} = \mathcal{I}_i^j
\frac{1 - exp [ - \bar{\sigma} n_{H,i}^{j+1} \, a_i^{j+1}]}{1 - exp [- \bar{\sigma} a_i^{j+1} n_{H,i}^j]}
\end{equation}

\subsection{The heating function $\gamma$}
The heating function in the thermal evolution is mainly provided by a) the photoelectric
emission of electrons from grains $\gamma_{pe}$ caused by the incident
FUV component ($6.5 < h\nu < 13.6$ eV) of background star light, which
can be treated as isotropic and has been discussed by \citet{nelson}.
so we don't repeat them here; b) the hydrogen ionisation
heating $\gamma_{ion}$ produced by the illumination of
UV component ($h\nu > 13.6$eV) from a nearby massive star,
which
is highly directional and whose strength is dependent on the distance of the cloud from
the star. Its formula can be
derived through the following considerations.
                                                                                                                              
When an atom with ionisation energy $E_1 $ is ionised following the absorption of one photon of frequency
$\nu$ and energy $E = h \nu$, it releases one electron which carries the excess
kinetic energy $ E_{\nu} - E_1$ which will be transfered to the gas through collisions with
other gas particles. The resulting heating rate is described by the following equation.
\begin{equation}
\gamma_{ion}  =  n (1-x) \sigma J k T_*
\end{equation}
where $k$ is the Boltzmann constant and $T_*$ is expressed as \citep{canto}
\begin{equation}
T_*=T_{eff}\frac{x_0^2 + 4 x_0 + 6}{x_0^2 + 2 x_0 + 2}
\end{equation}
where $T_{eff}$ is the stellar temperature, and $x_0 = (h\nu_0)/(k T_{eff})$ with $\nu_0$ being the
frequency of the Lyman limit.
                                                                                                                              
Heating of the gas is also affected by cosmic ray heating,
H$_2$ formation heating and gas-dust thermal exchange, for which the same
formulae as that in \citep{nelson} are used in our model.

\subsection{The cooling function $\Lambda$}
The cooling function $\Lambda$ is affected by recombination of the electrons with
ions, the collisional excitation of OII lines  and by CO, CI, CII and OI line emission.
                                                                                                                              
When a free electron in the plasma is captured by a proton, a photon is emitted and an amount of energy
$E_1 + m_e v_e^2 / 2$ is removed from the internal energy of the gas. The cooling rate due to this
process is \citep{hummer}
\begin{equation}
\Lambda_{rec} = \beta_B \, n^2 x^2 k T
\end{equation}
where $\beta_B = \alpha_B \times (1 + 0.158 t_e)$.
                                                                                                                              
Another important cooling resource comes from the collisional excitation of low-lying energy levels of OII
ions in spite of their low abundance, since OII has energy levels with
excitation potentials of the order of $kT$.  We use the following simplified formula derived by \citet{raga}
to calculate the cooling rate due to the collisional excitation of OII,
\begin{equation}
\Lambda_{colli} = \Lambda_{colli \, (1)} + \Lambda_{colli \, (2)}
\end{equation}
with
\begin{equation}
log_{10}[\frac{\Lambda_{colli \, (1)}}{n_e n_{OII}}] = 7.9 t_1 - 26.8
\end{equation}
\begin{equation}
log_{10}[\frac{\Lambda_{colli \, (2)}}{n_e n_{OII}}] = 1.9 \frac{t_2}{|t_2|^{0.5}} - 20.5
\end{equation}
with $t_1$ = $1 - 2000$ K$/T$ and $t_2 = 1 - 5 \times 10^4$ K$/T$ and $n_{OII} = x n_{OI}$.
The above formula is valid under the limit of low electron density $n_e < 10^4$ cm$^{-3}$, which
is true for the BRCs' structure.
                                                                                                                              
The cooling affects $\Lambda_{line}$ by CO, CI, CII and OI line emission are same as those in the
\citet{nelson}.
\subsection{Boundary conditions and initial velocity field}
The molecule cloud is assumed to  have a spherical symmetry and its mass
is uniformly distributed through a sphere of radius $R$. It is embedded in surrounding hot medium with a constant 
density (HII region) \citep{Kesselb}. The external pressure by the surrounding hot medium provides an inwardly directed
pressure force whose characteristic magnitude is similar to that expected from a HII region.  
The value of the assumed boundary pressure $P_{ext}$ we adopted here is equivalent to an ionised gas with
$n_e = 100 $cm$^{-3}$ and $T = 10^4$K\citep{lefloch2}, which resembles to 
Lefloch's zero radial velocity boundary condition.   

An outflow condition is imposed at a fixed radius equal to several times 
of the original radius of the cloud. Material that expand beyond this radius is simply removed 
from the calculation. By this time, these particles represent very low density material and so provide little UV 
extinction or other influence in the simulations. 
                                                                                                                              
The initial turbulent velocity field was set up in the usual manner, as described by \citet{maclows},
with perturbations set up with a flat power spectrum $P(k)$, and with a minimum wavenumber $k_{min}$
corresponding to a wavelength equal to the cloud diameter. Eight wave numbers were used to set up
the velocity perturbations with the maximum, $k_{max}=8 \, k_{min}$. The velocity perturbations were drawn
from a Gaussian random field determined by its power spectrum in Fourier space.
For each three dimensional wavenumber $k$, we randomly select an amplitude from a Gaussian distribution
centred on zero, and of height $P(k)$, and selected a random phase uniformly distributed between 0 and
$2 \pi$. The field is then transformed back into real space to get the real velocity for each particle,
and then multiplied by the amplitude needed to obtain the required total kinetic energy for the cloud.
This procedure is repeated for each velocity component independently to get the full three dimensional velocity
field. The amplitude of the turbulent motion was normalised such that the total kinetic energy
of the clouds equals their initial gravitational potential energy.

The initial geometry of the cloud and
radiation field configurations are shown in Figure \ref{geometric}.
Since the distance of the cloud to the nearby illuminating star is assumed to be large compared 
to the radius of the cloud,
the UV radiations from the star can then be treated as a plane-parallel radiation
onto the front surface of the cloud \citep{lefloch,Kessela,williams} while the FUV 
radiation is isotropic to the cloud \citep{gorti, nelson}

The radiation flux distribution at the boundary is
                                                                                                                              
\[  J ( r = R, \theta ) = \left\{ \begin{array}{ll}
   J_0(\mbox{Lyman}) + J_0(\mbox{FUV}) &  \, 0^o \le \theta \le 90^o) \\
   J_0(\mbox{FUV})  &  \, 90^o \le \theta \le 180^o)
                      \end{array} \right. \]
where $J_0(\mbox{Lyman}) $ is the value of the Lyman continuum flux at the surface of the cloud for which we
take the typical value of $2 \times 10^{11}$ cm$^{-2}$ s$^{-1}$ \citep{miao,williams};
$J_0(\mbox{FUV})$ is the value of the interstellar radiation at
the surface of the cloud which has the value of 100$\times G_0$, where $G_0$
 is the mean interstellar medium radiation flux and 
traditionally known as the 'Habing flux' \citep{habing};  $\theta$ is the azimuthal
angle in a spherical coordinate. 
\subsection{Classification of the initial condition of a cloud} 
\citet{lefloch} classified the initial physical properties of a cloud and UV radiation field in 
a two dimensional parameter space ($\Delta, \Gamma$), which are defined by  
the following equations\citep{lefloch}, 
\begin{eqnarray}
\Delta = \frac{n_i}{n_0} = \frac{c_i}{2 \eta \alpha_B n_0 R} 
\{ - 1 + \sqrt{ 1 + \frac{4 \eta \alpha_B R J_0(Lyman)}{c_i^2}} \}  
\label{parameter1}\\
\Gamma = \frac{\eta \alpha_B n_i R}{c_i} = \frac{1}{2} \{ - 1 + \sqrt{ 1 + \frac{4 \eta \alpha_B R J_0(Lyman)}{c_i^2}} \}
\label{parameter2}
\end{eqnarray}
where $\Delta$ relates the overpressure of the ionised gas to the undisturbed neutral gas  
and $\Gamma$ is the ratio of recombination rate to the ionisation rate,  $n_i = x n = n_e$ is 
the ionised gas density around the cloud, $n_0$ is the initial density of 
the hydrogen in the cloud.    
$\eta \approx 0.2 $ is a parameter to describe the effective thickness of the recombination layer
around the molecular cloud and $c_i = \sqrt{{\cal{R}} T_* }$ is the isothermal sound speed in the ionised
gas with $\cal R $ being the gas constant. 
The 2-D parameter space 
can be divided into five different regions according to the effect of the UV radiation on the 
 evolution of a cloud.  

Region I is defined by $\frac{\Gamma}{\Delta} \le 1.4\times10^{-2}$, where the opacity of the cloud 
is very low ( $\ll 1$ ) so that it is initially transparent to the UV radiation.
Region II is defined by $\Delta \le \frac{c_n^2}{2c^2_i} \sim  10^{-2}-10^{-3}$ with $c_n= \sqrt{{\cal{R}} T_n}$ 
being the isothermal sound 
speed in the neutral gas of temperature $T_n$, in which the effect of the ionisation is too 
weak to produce any noticeable dynamical
effect. Therefore clouds starting in region I and II are too trivial to be discussed.

On the other end, when $\Delta$ is high, i.e., 
$2 \le \Delta \le \sqrt{10}$, which  defines the region III, 
the cloud is entirely photo-ionised in an ionisation flash by Lefloch's model in which the 
self-gravity of the cloud is not taken into account.
Region IV is defined by $\Delta < \frac{J_0(Lyman)}{n_0 c_i (1 + \Gamma)}$, where the whole dynamical
evolution of the cloud will be governed by the propagation of a D type ionisation front preceded by an 
isothermal shock in the cloud. Finally $\Delta > \frac{J_0(Lyman)}{n_0 c_i (1 + \Gamma)}$ defines 
the region V, where an initial weak R ionisation front 
would gradually become a D type ionisation front propagating toward the interior of the cloud.    

For the clouds starting in regions III - V, although the detailed evolutionary features of 
the cloud can only be obtained  
through the numerical simulations described above, 
an existing simplified analytical result derived by \citet{kahn} could  
help astronomers predict the final  evolution of the globule under observation
and also provides us a way to validate our numerical results. Therefore we briefly present it in the following
subsection. 
\section{A preview: A simplified analytical result}
In the simplified analytical result, only the effect of UV  component is considered
since it is much stronger than FUV radiation as given in section 2.5. The viscosity term 
in Equation (2) will be neglected because of its smaller magnitude than the other terms.

After an initial ionising stage by the UV radiation flux, 
the stream of the ionised hydrogen
forms a warm halo around the globule which has a radius $r_i$, 
at which the globule is bounded by an ionisation front while an isothermal shock
propagates toward the interior of the globule. Assuming a steady ionised gas
flow from the surface of the globule which can be approximated as a spherical isothermal sphere and 
a balanced ionisation-recombination in the halo, 
\citet{kahn} derived a necessary condition for the globule to be stable against the gravitational collapse.
Based on the analytical solutions of Equation (1-2) in the
ionised gas halo and in the neutral globule, this critical condition  
is expressed by the current mass of the neutral globule 
\begin{equation}
M \le M_{max} = 1.29( \frac{\alpha_{B} c_n^{14}}{16 \pi^2 m_H^2 c_i^4 J_0({\mbox Lyman})G^5})^{1/3}
\label{mass1}
\end{equation}
where $G$ is the gravitational constant. 

The importance of this result is that 
the quantities such as $T_n(c_n)$, $T_i(c_i)$ and $J_0$(Lyman)  
 can be estimated by the observational data, then
Equation(\ref{mass1}) can be used to estimate the maximum mass for the globule to contain
in order not to go collapse if the structures of both the globule and halo
can be regarded as steady. On the other hand if the current mass $M$ of the globule
derived from the observational   
data is higher than $M_{max}$, the globule under the investigation would  
finally collapse to form new star at the end of the evolution, vice verse, the globule may stay stable 
until it is completely evaporated
by the UV radiation. The advantage of the 
 Equation(\ref{mass1}) is that the criteria for globule's stability is in terms of
 the current mass of the globule instead of the initial mass of the cloud, 
 the latter is hardly known by observations.

In the following, we will discuss the detailed evolutionary features of clouds,
based on the numerical simulations.  In this paper we will mainly 
study the morphological evolution of BRCs and try to find solutions 
for the questions raised in the beginning of the
paper. A systematic investigation
on the ionisation radiation triggered star formation in BRCs will be presented in our next
paper. 
\section{Results and discussions}
The most important difference between
our newly developed RDI model and Lefloch's model is the inclusion of the self-gravity
of a molecular cloud, therefore we will firstly investigate the role of  the self-gravity of a molecular
cloud in its dynamical evolution when it is under the effect of UV radiations, so that we can 
explore the possibility to relax the extreme zero velocity boundary condition
in order to derive the predominance of type A BRCs, and discuss the possible modification to the 
classification of Lefloch's 2-D parameter space caused by the inclusion of the self-gravity of 
the molecular cloud.
     
We then will explore the effects of initial thermal state of a cloud, 
the strength of UV radiation flux, and also initial turbulent state
(non-thermal motion) on the morphological evolution of molecular clouds.
                                                    
The molecular cloud is presented by 20,000 SPH particles. In the following analysis to the numerical results, 
we will concentrate on the effect of the UV radiation falling onto the surface of the front hemisphere, 
since the heating effect of photoelectric emission of electrons from dust grains by the isotropic 
FUV radiation is much weaker than that of hydrogen ionisation by Lyman continuum radiation 
which only affects those gas particles at the front surface layer.
  
\subsection{The effect of initial self-gravity  of a molecular cloud}
The gravitational potential energy of
 a spherically uniform molecular cloud of mass $M$ and radius $R$   
is $\Omega = - \frac{3 G M^2}{5R}$. In order to
explore the role of the initial self-gravity of a molecular cloud in its morphological evolution,
  we perform two different sets of simulations, one set of the three clouds having 
a same mass of 35M$_{\sun}$ but different radii, the other having a same radius of 
0.5 pc but different masses. Therefore our simulations cover different cases in which 
the initial self-gravity of a molecular cloud changes. 

The initial temperature of the molecular clouds is all set to 60 K. 
These chosen initial conditions
guarantee that the simulated molecular cloud will not collapse by  self-gravity if there is no
external radiations. 
\subsubsection{Clouds with a fixed mass of 35 M$_{\sun}$}
Due to the change of the radii of clouds, the initial ionisation/recombination states change 
as well. The first six columns in Table 1. list  the parameters of the first set clouds called A, B, C 
 and their initial locations in the ( $\Delta, \Gamma$ ) 
parameter space respectively. The features of cloud D in the list will be discussed later.
 The ratio of the initial gravitational energy of the three cloud A, B, and C 
is $\Omega_A : \Omega_B : \Omega_C = 1 : 0.38 :0.26$ and the  ratio of 
gravitational forces of the clouds A, B and C onto a mass element at their 
surfaces are correspondingly 
$F_A : F_B : F_C = 1 : 0.14 : 0.07$, which  defines 
significantly different initial self-gravity in the three  molecular clouds. 

According to Lefloch's model, cloud A would evolve into a 
cometary globule after passing a morphological sequence from 
type A to type B then to type C, and finally it will be completely 
evaporated in a few Mys, while cloud B and C
would be evaporated in a so called 'ionisation flash'. It is our interest to see how these clouds 
would evolve differently if the self-gravity of a cloud is taken into account. 

\noindent{\bf I: CLOUD A}

The images in Figure \ref{figa1} 
reveal the morphological evolution of cloud A by number density of hydrogen nuclei
  and  the line plots in 
Figure \ref{figb1} describe how the UV radiation induced shock front leads
the ionisation front propagating into the interior of the cloud and finally 
results in the collapse of the cloud. Figure \ref{figc1} is the corresponding
velocity field evolution diagram, which gives the kinematic explanation on the 
formation of the morphologies of the cloud at different evolutionary stages.

The image in the top-left panel of Figure \ref{figa1} and the solid line in the
top-left panel of Figure \ref{figb1} show the number density distribution of cloud A
at an early stage ($t = 700$ years). It is seen that the density profile of the whole cloud 
has not yet changed much from the initial uniform distribution. The top-left panel in Figure \ref{figc1}
shows a rather random velocity distribution which is very  similar to the initial random 
velocity distribution of the cloud.   

However the dashed line in the top-left panel of Figure \ref{figb1} tells that 
an ionisation front has been built up at the front surface of the upper hemisphere 
of the cloud after the intensive Lyman continuum flux falls onto the front surface.
The gas in the top layer is fully ionised ($x \sim 1$) and the resulted ionisation heating rises
its temperature to $10^4$K. When $t = 0.2 \times 10^5$ years, the
ionised gas at the front surface of the cloud evaporate radially away from the front 
surface as shown by the velocity profile in the middle panel of Figure \ref{figc1} 
and by the furry gas above the  
front surface in the top-middle panel of Figure \ref{figa1}. The falling tail beyond $z =  0.4$pc
in the solid line of top-middle panel in Figure \ref{figb1} shows that the density of the evaporated 
gas is about 100 cm$^{-3}$.  
  
On the other hand, the high temperature in the front top layer 
produces a high pressure, i.e.,  an isothermal shock driven into
the cloud. The neutral gas ahead the shock front is then compressed. The contour lines around
the front surface layer of the cloud in the top-middle panel in Figure \ref{figa1} and the small 
peak (if we define it as the shock front) around $z = 0.4$pc in the solid  
line of the top-middle panel in Figure \ref{figb1} show that a density
gradient starts to built up  as a consequence of the shock compression. 
                                                                                                                              
The UV radiation induced shock velocity $v_c(\theta)$ at the front surface 
 is approximately given by \citet{lefloch},
\begin{equation}
V_c(\theta)= V_c(0)(cos(\theta))^{1/4}
\label{velo-e}
\end{equation}
where $\theta$ is the angle of a surface point from $z$ axis and $ 0
\le \theta \le \pi / 2 $ as shown in Figure \ref{geometric};  $V_c(0)$ is the shock velocity 
at the the point $r=R,\theta = 0 $ and $V_c(0) \sim [J_0(\mbox{Lyman}) / n ]^{1/2}$ \citep{bertoldi}.
The direction of $V_c$ is along the normal of the point (R,$\theta$). The resulted velocity 
distribution $V_c(\theta)$ in the shocked gas is obviously seen in the top-middle panel of Figure \ref{figc1}.

At time $t =2.0 \times 10^5$ years, the shock front leads the ionisation
front moving toward the rear hemisphere as shown by the solid and dashed lines 
in the top-right panel of Figure \ref{figb1}. The ionisation ratio $x$ has decreased a lot in 
the shocked part of the cloud, because
the recombinations of electrons with their parent ions in the surrounding envelop  
have consumed most of Lyman photons.   
 On the other hand the shocked gas at the front surface 
around $\theta \sim 0$ have the highest $V_c$ so it moves faster toward the
rear hemisphere than that in the peripheral part of the compressed layer.
 Consequently, the faster progression of the central parts of the compressed layer
at the front surface causes a deviation of
the morphology of the front surface  from a hemisphere, i.e, the front
surface of the hemisphere becomes seriously squashed so that a type A rim  forms,
which can been seen from the top-right and the subsequent panels of Figure \ref{figa1}.
                                                                                                                              
At the same time, the strongest compression of the gas
located at $\theta \sim  0 $ 
forms a small but highly condensed core at the head of the
front hemisphere, shown as in the
top-right panels of Figures \ref{figa1} and \ref{figb1}, which creates 
a new gravitational centre G to its surrounding gas particles so that the latter gain 
 an acceleration $a_G$ hence a velocity component $V_G$ as shown in   
Figure \ref{figv}. The surrounding gas then
gradually collapse toward the core centred at G, as shown  
in the panels from the top-right to the bottom-right panels in Figures \ref{figb1} and 
\ref{figc1}. 

Due to the distortion of the front surface, 
the velocity of the shocked gas in the outermost parts of 
the front hemisphere ( $\theta \sim \pi / 2$ )
is significantly changed so that its trajectory
intersects the rear edge of the original cloud 
to form the 'ear-like' structures as shown by the two
convex parts at $\theta \sim \pi/2 $ 
in the top-right and bottom-left panels of Figure \ref{figa1}.   

As illustrated in Figure \ref{figv}, the total velocity $V_T$ of the gas particle 
at the sample point in the 'ear-like' structure directs to the C' point on the $z$ axis, 
instead of C point, which would be the converging point if the self-gravity of 
the globule was not taken into account. Hence  
the 'ear-like' structures converges to $z$ axis to form a tail structure which overshoots
the rear surface of the cloud  and is well described by the solid line
beyond $z < - 0.5$pc from the top-right to bottom-right panels of Figures \ref{figb1} and
\ref{figc1}. 

At $t$ = 0.351 My, the type A BRC structure comes to a thermally quasi-static state
(with a pressure $P/k \sim 10^6$ Kcm$^{-3}$ close to the boundary between the neutral
gas and the ionised envelop)  after an initial drastic
and transient evolutionary period.  As shown in Figure \ref{figb1}, the whole structure has 
shown the typical stratification of an observed BRC \citep{lefloch2} from the centre to outside:

i)  An isothermal neutral core of a radius of 0.15 pc ($ - 0.3 < z < 0$ pc) with a central density 
of a few $10^5$ cm$^{-3}$ and an average temperature of 
$T_n$ = 28 K

ii) A very thin layer of PDR 
(Photon-Dominated Region: $0 < z < 0.05$ pc) with 
 an average density $ 10^4$ cm$^{-3}$, an temperature $\sim 1000$ K and an  
electron density $n_e < 10$ cm$^{-3}$ with the ionisation fraction $10^{-5} < x < 10^{-3}$. 

The mass included in the isothermal core including the PDR (with density from $10^4$ 
to a few $10^{5}$ cm$^{-3}$ as observed) is 15 M$_{\sun}$.   

iii) A  photoionised envelope (Bright rim: $0.05 < z < 0.11$pc),
the density drops sharply from a few $10^3$ to a few 10 cm$^{-3}$  
and the ionisation ratio $ 10^{-3}< x \le 1$, i.e.,  $n_e$ distribution between
10 and $10^3$cm$^{-3}$. The average temperature $T_i$ is about $10^3$ K.

iv) The evaporated ionised tenuous gas into the HII region ($z > 0.11$pc) 
 with an average density a few 10 cm$^{-3}$ and $x = 1$, but a very high average 
temperature of $5 \times 10^4$ K.  

Now the most concerned 
question on its further evolution is that whether the whole globule is stable against the 
gravitational collapse or not. Before revealing the numerical solution to this question, 
we can employ Kahn's analytical result to predict this globule's  final 
evolution and then use the result to validate the numerical solution. 
Substituting all of the values
into Equation (\ref{mass1}), we obtain the maximum mass for the isothermal core to be stable is 
M$_{max}$ = 3 M$_{\sun}$. Therefore this type A BRC should finally collapse since  
the current mass of the core is 15 M$_{\sun}$, much higher than M$_{max}$.  

The numerical results after $t$ = 0.35 My as shown by the bottom middle/right panels in 
Figures \ref{figa1}, \ref{figb1} and \ref{figc1} reveal that the globule quickly 
collapse toward its G centre under the effect of the enhanced self-gravity. 
The central density of the core
reached 2$\times 10^{8}$cm$^{-3}$ when $t$ = 0.391My and 5$\times 10^{10}$ cm$^{-3}$ when $t$ = 0.394My.
Therefore we could conclude that  the radiation triggered star 
formation in cloud A is expected to occur because the central density has
reached such a high value \citep{nelson}, which is consistent with the analytical result. 

\noindent{\bf II: CLOUD B}

The morphological evolution of cloud B is shown in
Figure \ref{figa2}, while  Figure \ref{figb2} describes how the shock front leads
the ionisation front propagating into the cloud and Figure \ref{figc2} displays the
corresponding evolutionary sequence of its velocity field.

During the first $0.4$ My, the main characteristics of the morphological
evolution of cloud B are similar to that
of the cloud A.  The gas particles at the top layer of the front hemisphere are continuously
 ionised, evaporated and form a hot photo-ionised envelope around the type A rim at the 
front surface of the globule. 
On the other hand, the radiation induced shock leading an ionisation front propagates 
into the cloud and compresses the 
neutral gas ahead to form a condensed core. 

However the morphological evolution of cloud B starts to show
obviously different feature at $ t = 0.63 $My. It is seen from the bottom-left
panel of Figures \ref{figa2}  that  the front surface becomes a type B rim with
 a dense core behind the rim. Also a more extensive tail than that in cloud A is 
formed by the gas particles from the 'ear-like' structure
converging to $z$ axis as shown by the solid lines beyond $z < - 1$ pc in the bottom panels 
of Figure \ref{figb2}. We will discuss the kinematics which make the difference in
the morphological evolutions between cloud A and cloud B after the presentation of
the evolutionary features of cloud C.  

Around the front hemisphere of the type B BRC, a similar thermally quasi-stable stratification 
to that of cloud A at $t = 0.351$ My is formed with an isothermal  
condensed core of a radius of 0.2 pc at the
head of the globule  as shown in the bottom-left panel of 
Figure \ref{figb2}. The core has a central density slightly lower than $10^5$ cm$^{-3}$, 
and an average temperature $T_n = 31$ K, which is surrounded by a thin layer of 
photo-ionised envelope of a temperature $T_i = 10^3$ K. The core including the PDR contains 
a mass about 13.5 M$_{\sun}$ which is higher than the $M_{max}$ =3.6 M$_{\sun}$ given by  
Equation(\ref{mass1}).
Therefore the core should 
be expected to collapse. The simulation results presented in the bottom middle/right panels of Figure
 \ref{figa2} and
Figure \ref{figb2} show that the core continuously gets denser, and the central density has reached
a few $\times 10^5$ cm$^{-3}$ at $t$ = 0.79 My and a few $\times 10^{10}$ cm$^{-3}$ at $t$ = 0.81 My, the
sign of the triggered star formation which  
is consistent with the analytical prediction.   

\noindent{\bf III: CLOUD C}

The evolutionary sequences of density, ionisation ratio and the velocity field of cloud C 
are shown in Figures \ref{figa3}, \ref{figb3} and \ref{figc3} respectively. 
In the first 0.86 My, the dynamical evolution of the cloud C  
displays similar features to that of cloud B. A type A rim  
firstly forms after $t = 0.365$ My, and then
a type B rimmed morphology develops when $t = 0.86 $ My, 
as shown in the corresponding panel in Figure \ref{figa3}. The relevant
panel in Figure \ref{figb3}  shows
the formation of an isothermal core with a 
central density slightly lower than 
$10^5$ cm$^{-3}$ and a thin layer of PDR surrounding the core. 
The average temperature $T_n$ is 35 K in the core 
and $T_i = 1000$ in the photo-ionised bright layer. The thermal pressure 
$P/k$ in the different regions is $\sim 10^6$. The mass included within the isothermal 
core (PDR included) is 
11.1 M$_{\sun}$, which is higher than the 5 M$_{\sun}$ estimated by Equation (\ref{mass1})      
Therefore, the core would collapse finally. 

However, the gas particles at the two sides of the rear hemisphere continue their movement  
to $z$ axis to form a longer tail as shown in the bottom-middle panels of Figures \ref{figa3}
and \ref{figc3} so that the morphology of the globule becomes a type C BRC at $t = 1.034$ My.  

The last two panels in Figure \ref{figb3} shows that 
the gas particles in the core region collapse to the $G$ centre, so the central density reaches 
to $10^9$ cm$^{-3}$, which means the beginning of the triggered star formation at $t = 1.12$ My.   
                                                        
In table 1. the final  morphologies of the simulated clouds are list in the 7th column 
and the 8th and 9th columns are the isothermal Core Formation Time (C.F.T.)
and the Core Collapse Time(C.C.T.) respectively. 

In order to find the mechanism responsible for the differences in the 
morphological evolution of cloud A, B and C, a further investigation 
on the kinematics of the shocked gas in the front hemisphere is carried in the following part of 
the paper.

\noindent{\bf IV: The Kinematics of morphological evolution}

Figure \ref{figv} shows that the direction of the total velocity $V_T$ of the shocked gas particles 
within two 'ear-like' structures determines their converging points on the $z$ axis, hence 
the length of the tail or say it is the magnitudes of $V_C$ and $V_G$ which 
determine the morphological evolution of a BRC.   

The ratio of the gravitational accelerations of cloud A, B and C 
on the gas particles in the 'ear-like' structure is $a_A : a_B :a_C = 1:0.14:0.07$. 
 Hence the relevant velocity components obey the inequality 
$V_G(A) > V_G(B) > V_G(C)$.
On the other hand, as stated in Equation(20), we have $V_C(\theta)\propto V_C(0) \sim n^{- 1/2}$ for
fixed $J_0$ and $\theta$,  
hence for the particles in the the 'ear-like' structures in cloud A, B and C, the velocity component
$V_C$ follows $V_C(A) < V_C(B) < V_C(C)$. 
Now it is not difficult to understand that the angle $\delta$ of a gas particle in the 'ear-like'
structure in  cloud A, B, C obeys the inequality: $\delta(A) > \delta(B) > \delta (C)$.
The distance $d$ between the converging point C' and the new gravitational centre G
in clouds A, B and C obeys  $d_A < d_B < d_C$, therefore the three clouds
evolved to type A, B, and C BRCs respectively.  

On the other hand, the escape velocity of a particle at the surface of a cloud is 
$v_{es}= \sqrt{\frac{2GM}{R}}$. The ratio of the escaping velocities 
of three clouds is $v_{A(es)} : v_{B(es)}: v_{C(es)} = 1:0.37:0.26 $. The easiness 
for the mass particles at the surface of the cloud to be photo-evaporated is in the order
of cloud C, B, and A, which explains why the masses left in the isothermal cores are 
11.1, 13.5 and 15 M$_{\sun}$ for clouds C, B, and A respectively.  

\noindent{\bf V: Further increment of the radius}   

We further increase the initial radius of the molecular cloud so that the cloud is initially much
less gravitationally bounded than cloud C. The basic features in the evolutionary sequences of 
the molecular clouds are not much different from that of the cloud C if $ R < 4.7$ pc except that
both the mass accumulated in the head of the quasi-stable globules and the central density decrease with
the initial radius increasing from 1.9 to  4.7 pc.

However when the initial radius increases to $R = 4.7 $ pc (the cloud is called cloud D in Table 1), the
initial gravitational force to the surrounding gas particles is only about 1.1\% of that 
of cloud A.    
Cloud D quickly evolves from type A to B then into a loose type C BRC -- a cometary globule with 
a slightly compressed core of a central density of a few $10^2 $ cm$^{-3}$ when $t = 0.8$ My.
Shortly after a weak shock passed the centre 
of the head, the whole globule undergoes a quick
re-expansion and is totally photoevaporated after 1 My in 
a so called 'ionising flush' \citep{lefloch}. 

The above set of simulations prove that changing the initial size of the cloud hence the initial
self-gravity could dramatically change the cloud's evolutionary sequence and the final morphology. The
result means the initial self-gravity of the cloud plays a dominant role in the 
morphological evolutions of BRCs. However, changing of the initial self-gravity of the cloud 
by changing its initial radius creates a second effect at the same time - the alteration of 
its initial ionisation/recombination state, as shown in Table 1. 
Hence our confidence in the role played by the self-gravity in determining 
the morphological evolution of BRCs needs further assurance. The second set of simulations 
therefore is designed to change the initial density of the cloud by changing its initial mass
so that the change in the initial self-gravity will not bring drastic change 
in the ionisation/recombination state as shown in Table 2.
 Hence the role of the self-gravity in the morphological
evolution of the BRCs can be further explored.     
\subsubsection{Clouds with a fixed radius of 0.5pc}
Table 2. lists the parameters of the second set clouds called A, A$^{'}$, A$^{''}$ 
respectively with their initial locations in ( $\Delta, \Gamma$ ) 
parameter space and the information on their morphological evolutions.
 
For cloud A, the evolutionary features has been fully discussed before. 
and are re-listed in Table 2 for comparison.
With the decrement of the initial mass of the cloud, the density of the cloud 
decreases accordingly, the morphological evolution follows  a similar 
pattern to that in the first set of clouds. The dynamical evolutions of the formation
of type A, B and C BRCs exhibit similar sequences to that in the first set clouds. Therefore 
in order to save space, we don't present the sequential figures on their dynamical 
evolutions as we did for set one clouds, but only provide a summary on the relevant key features 
which are partly listed in the Table 2.
 
When the initial mass of a cloud decreased to 12 M$_{\sun}$, cloud A$^{'}$ evolved from type A to 
B morphology in a very similar way to cloud B in set one simulation. An isothermal
core formed in 0.42 My, and collapsed in 0.8 My, with the central density up 
to $10^8$cm$^{-3}$ and a final mass of 6.9 M$_{\sun}$ in a collapsed core of 0.1 pc in radius. 
When the initial
mass  further decreased to 8 M$_{\sun}$, cloud A$^{''}$ evolved through
a sequence of from type A to B then to C morphologies, with formation of an isothermal core
at $t$ = 0.53 My, which collapsed in 1.5 My with a central density of $10^7$cm$^{-3}$
and a mass of 4.5 M$_{\sun}$ in a core of radius of 0.1 pc. 

Finally when the initial cloud mass further decreased to 1 M$_{\sun}$, cloud A$^{'''}$ 
evolved quickly through type A and B morphologies with a condensed core of 
a central density of $10^4$ cm$^{-3}$. After evolving to a cometary globule morphology at t = 0.3 My, 
it re-expands and then is totally evaporated when t = 0.9 My. 
   
In the second set simulations, the initial recombination parameter $\Gamma$ is  same for  
 cloud A, A$^{'}$ and A$^{''}$  because they have same radius.  
The ratio of the initial ionisation parameters $\Delta(A):\Delta(A'):\Delta(A'') = 1 : 2.85 : 4 $,
i.e., cloud A$^{''}$ has the highest value of $\Delta$ which is  4 times the lowest value for cloud A. 
On the other hand, the ratio of the initial 
gravitational energies of the three clouds   
is $\Omega_A : \Omega_{A'} : \Omega_{A''} = 1 : 0.12 : 0.052$. The highest value of the initial 
self-gravitational energy of cloud A is about 20 times of the lowest value for cloud A$^{''}$. It is obvious 
the difference in the initial self-gravitational energies among the three clouds is much more distinctive than 
the difference in their initial ionisation states. Therefore, we can safely conclude that the different 
morphological evolutions of the three clouds are caused by their different initial self-gravitational states.  
The role of the initial self-gravity in determining the morphological evolutions of the BRCs
 is further confirmed.   

\subsubsection{The Possible modification of the ($\Delta,\Gamma$) parameter space}
From the morphological evolution paths of the 7 clouds, we see that for clouds starting from region V 
such as cloud A and A', their dynamical evolutions are driven initially by a  weak-R ionisation
front which quickly changes into a D-type ionisation front, just as what Lefloch described. 

However for clouds starting from region III such as cloud B,C,and A'', they
are not at all entirely photo-ionised by an R-weak front and permeated by an ionisation flash like 
what \citet{lefloch} described, their dynamical evolutions are driven by a similar mechanism 
to that of cloud A and A'.  

For clouds A$^{'''}$ and D, which also start from region III but with a high initial ionisation
state $\Delta \ge 23$, its dynamical evolution is 
driven by a weak R-front and the whole structure get permeated by an ionisation flash 
and evaporated quickly just as what \citet{lefloch} described for region III cloud.   

Hence it is clear that in term of the dynamical classification
of ($\Delta, \Gamma$) parameter space,  the boundary for region III has been pushed greatly
from $\Delta < \sqrt{10}$ to  $\Delta < 23$ due to the inclusion of the self-gravity of the cloud. 

\subsubsection{Zero self-gravity and surface instability of the cloud}
In order to further support our argument on the role of self-gravity of the cloud in the morphologic evolution 
process, we take off the self-gravity of the three clouds A, B and C  and re-run the simulations. The result 
reveals very similar evolutionary sequence to that of \citet{lefloch}. The evolutionary sequence passes
type A to B morphologies and then form a type C BRC. Then it re-expands
and finally totally evaporated between 1 to 2 My. 

To some stage of the evolution of the molecular cloud, the ionisation flux onto the front surface of 
the cloud is perturbed somehow due to the recombinations of the electrons onto the ionic atoms, 
 which causes 
instability of the front surface of the cloud as described in the lefloch's simulations \citep{lefloch}.
We found from our simulations that the scale of the instability is closely related to the
initial ionisation state of the cloud. For all of the 7 clouds we simulated, except cloud D, their initial
ionisation state are not very high when compared to that of the cloud D, very small scale perturbations
(with wavelength $\lambda < 0.2 R << R)$) at the front surface appeared, as seen from the up-middle panels
of Fig.2, 6 and 9, but they are quickly smoothed out.    

For cloud D, due to its  extremely high initial ionisation state, 
the surrounding of the front surface quickly 
becomes recombination dominant region, so large scale perturbations (with wavelength 
$0.2 R < \lambda < R $) grow at $t = 0.06$ My from the front surface of the cloud which 
can be seen from the up-left panel of Figure \ref{instability}. However the large scale 
perturbation is stable because it is gradually smoothed out with the core further being compressed by
the isothermal shock. At $t = 0.8$ My, the front surface becomes very smoothed. 
We think it is the gradually increased gravity from the core underneath the front surface which
suppressed further growth of the surface instability.       

We conclude that with inclusion of the self-gravity in the model, 
both small and large scale surface instabilities
caused by the perturbation to the ionisation flux are stable and 
will be smoothed out over the further evolution of the cloud. Our conclusion 
is in agreement with what revealed
by \citet{Kesselb} in Case B in their paper on the stability of the front surface of the cloud.

We have shown that the self-gravity indeed plays an important role in the morphological evolution of the 
molecular clouds. For any change in the initial conditions of a cloud, if it greatly 
results in a change in its initial self-gravitational state, 
the cloud will alter its morphological evolution.          

\subsection{Initially Thermally Supported Clouds}
In order to see how the three clouds would evolve if they are only initially thermally supported, i.e., if
the cloud A, B and C start their evolutions without the initial turbulence.  
We re-run the simulations for cloud A, B and C respectively by setting a zero initial velocity field to
each of them.   
Since the general features of the morphological evolutions of the three molecular clouds are 
qualitatively similar to that presented in last section, we mainly describe the different features in
their evolutions. 

Cloud A evolved into a type A BRC with a condensed core of a central density of a few 10$^{8}$cm$^{-3}$ 
within 0.37 My. Cloud B evolved into a type B BRC with a condensed core of a central
density 10$^9$cm$^{-3}$ within 0.74 My.   
Cloud C reached a type C BRC structure with a core density of $10^9$cm$^{-3}$ within 0.98 My. Compared with
the corresponding results in Section 4.1.1,  the clouds without initial turbulence take less time to collapse.
Further comparing the velocity field evolutions of the two group molecular clouds, 
we find the initially turbulent field takes 0.02 My in cloud A, 0.07 My in cloud B and 
0.14 My in cloud C to dissipate, which shows the initially turbulent clouds is just delayed to 
form a highly condensed core. Therefore we conclude that the initial turbulence does not effect a cloud's final
 morphology type. The results is 
consistent with that of an investigation by \citet{nelson} on the dynamical 
evolutions of Bok globules affected by FUV radiation.   
        
\subsection{Effect of changing initial density profiles}}
All of the above simulations are based on an initially gravitationally stable and uniform cloud. 
Although large degrees of central condensation should not be expected in the initially gravitationally stable
 molecular cloud,  we are still interested in examining the effect of the initial central condensation 
of a molecular cloud on its morphological evolution. The following initial density distribution 
is used, 

\[  n ( r ) = \left\{ \begin{array}{ll}
    n_0 \, \,  ( 0 \le r \le r_0) \\
   n_0 (\frac{r_0}{r})^2 \, \, (r_0 < r \le R)  
                      \end{array} \right. \]
which represents for a condensed uniform core with a radius of $r_0$ and density of $n_0$ surrounded
by an envelop of a density profile proportional to $r^{-2}$. For a given $r_0$, $R$ and $M$, $n_0$ can be found from
the equation $4\pi m_H \int_0^R n(r) r^2 dr = M$. 

When we start with $r_0 = 0.5R$, the evolutionary sequences of clouds A, B and C with an 
initial core-halo mass distribution
do not show significant change from that of initially uniform clouds. The basic features of 
type A, B and C are kept, except that the main body of the formed BRC structure are all smaller 
than that formed in initially uniform clouds. Then we decrease the radius of the central condensation to 
$r_0 = 0.2 R$, the morphological evolution show obvious changes. Cloud A ends its evolution in type B as shown
in Figure \ref{figa4}, and cloud B in  type C  as shown in Figure \ref{figa5}, 
while cloud C developed into a very narrow shaped type C at a much early time ( $t = 0.67$ My ) 
 as shown in Figure \ref{figa6}. 
Their morphologies are all elongated than that of their corresponding uniform  clouds.
This morphological change can be well explained by 
the illustration in Figure (\ref{figv}), the gas particles in the halo has a 
lower density therefore  a higher $V_C$ component, which makes them
 converge to a farther point C' from point G on the $z$ axis. 
  
The mass left at the collapsed core of 0.15 pc in cloud A,
B, and C are 10.5, 7.2, 5.6 M$_{\sun}$ respectively, which are less than that in their 
corresponding initially uniform clouds. This is because the density in the initial halo of a cloud is 
lower than that in the same part of the initially uniform cloud hence the halo in the front hemisphere is 
at a higher initial ionisation state, in which particles are easier to be ionised and then to be
photo-evaporated.        
 
In summary, the initial mass condensation state does affect a cloud's
morphological evolutions because it changes the initial self-gravitational state of a cloud.   
 
\subsection{Effects of the boundary conditions}
As stated in Section 2.5, we apply a constant pressure boundary condition, which  resembles 
the physical conditions of a very hot HII region.
In order to examine the effect of the boundary condition on the dynamical evolution of a cloud 
we decrease the value of  the surrounding pressure $P_{ext}/k$ to mimic a less violent environment.
For each of three clouds in the first set, we re-run the simulation for two different 
values of the external pressure, i) $P_{ext}/k = 10^3$ Kcm$^{-3}$ which corresponds to a warm interstellar medium 
environment \citep{nelson} and ii) $P_{ext}/k = 0$ which describes a vacuum environment so that the ionised gas   
is able to freely expand from the surface of the cloud. 

It is found that with decreasing external pressure, although cloud A still evolves to a type
A BRCs under the boundary condition i) and ii),  
the curvature of the front surface increases and the shape of the final condensed core becomes
elongated. Cloud B with boundary condition i) evolves to a slightly elongated type B BRC
but evolves to a type C BRC at $t = 1.6$ My under the boundary condition ii), as shown in Figure \ref{figa7}. 
Cloud C forms a type C BRC in both cases, but the length of the tail increases with the decrease of
the external pressure. It is obvious that the  
lower pressure environment favours formation of morphology with higher curvature, or 
high pressure environment favours formation of type A BRCs. The conclusion is 
consistent with that of zero velocity boundary condition \citep{lefloch}. 

\subsection{The effect of initial temperature of molecular clouds}
According to observations, the temperatures of molecular clouds are generally in the range of
$10 - 100$ K \citep{bertoldi}. We then carried out
two more sets of simulations for each of the three molecular clouds A, B and C
with the initial temperatures $T_1 = 20 $ K and $T_2 = 100$ K
respectively, while all of the others  physical conditions are kept the same as that in 
Section 4.1.1.
                                                                                                                                  
The evolutionary sequence of each simulation is qualitatively similar to
its corresponding cloud with an initial temperature of 60 K. 
                                                                                                                               
When the initial temperature $T_1 = 20$ K, the evolutionary sequences of cloud A, B and C
basically exhibit no significant differences from that
 described in Section 4.1.1. Cloud A evolves to a quasi-stable
type A rimmed morphology with a  condensed core at the head of
its front hemisphere. The central density of the core reaches
a value of $10^9$ cm$^{-3}$ in 0.37 My when it finally collapses. Cloud B evolves from
type A to B morphology with a condensed core at the head of
its front hemisphere in 0.6 My and the core reaches a central density of $10^{10}$ cm$^{-3}$
at $t = 0.79$ My. Similarly cloud C evolves to a type C BRC after 
passing type A and B morphologies. It collapse at $t = 1.01$ My when
the central density of the core is $2\times10^9$ cm$^{-3}$.

When the initial temperature rises to 100 K,
cloud A still evolves to a type A rimmed morphology having a condensed core
which collapses after the central density is up to a few $10^8$ cm$^{-3}$ in the front hemisphere
at $t =  0.42$ My. Cloud B evolves to a type B
rimmed morphology and the condensed core collapse with the central density being up
to $10^9$ cm$^{-3}$ at $t = 0.93$ My.
Cloud C evolves to a type C rimmed BRC plus a small core of
centre density of $10^9$ cm$^{-3}$ in 1.32 My. 

In summary, the variation of the initial temperature of molecular clouds between 10 - 100 K 
does not significantly change the morphological
evolution of a cloud, which is physically understandable 
since the thermal pressure of 
a cloud is not important in the dynamical evolution in BRCs \citep{bertoldi, lefloch, white}.
 Higher initial temperature in a cloud delays its collapsing time,
for the
initial thermal motions of gas particles dissipate over a period of  time. 
The morphological evolution of a cloud is mainly decided by the dynamical
behaviour of the gas particles within the two sides of the cloud ($\theta \sim \pi/2$), which is largely
dependent on the strength of the ionising flux and the self-gravity of the cloud as shown in 
Section 4.1.

However, the initial temperature of a cloud does affect the amount of  mass condensed in the core
of a BRC,  for which we will have a detailed discussion  when we study the UV radiation
triggered star formation in BRCs in our next paper.

\subsection{The effect of the strength of ionising flux}
We also carried out a set of simulations on the influence of the strength of an ionising flux on the
evolutionary process of BRCs. The change in the strength of ionising flux incident on the front surface
of a molecular cloud corresponds to two different cases: i) a cloud can be at different distances from an 
ionising star; ii) an  ionising star in different types radiates with different flux strength. 
 This set of simulations is to see whether the molecular clouds within one giant cloud cluster would 
exhibit essential difference in their morphological evolutions due to their different distances 
from the central star.  
    
Since the ionising flux $J(\mbox{Lyman}) \propto 1 / D^2 $, where $D$ is the distance of a cloud from the
ionising star. For each of cloud A, B and C, we re-run the simulations twice by setting two different 
$J(\mbox{Lyman})$ flux values. If we assume cloud A, B and C in Section 4.1.1 are in a distance of $D_0$ 
from the ionising star, then  $J_{1}(\mbox{Lyman}) = J_0(\mbox{Lyman})/ 4 $ 
 means we move the star twice distance away, i.e., 
$D = 2D_0$. If we set $J_{2}(\mbox{Lyman}) = 4 \times J_0(\mbox{Lyman})$
which means the original cloud is moved to half distance closer, i.e, 
 $D = D_0 / 2$. The spatial range  
of 0.5 - 2 $D_0$ from the ionising star should be large enough to include most molecular clouds 
in one cloud cluster.                                                                                                             

The simulation results revealed that for all of the three 
molecular cloud A, B, and C,  
their morphological evolutions are not
very sensitive to the strength of the ionising flux in the above specified range and
they follow very similar evolutionary sequences to that when ionising flux is $J_0(Lyman)$.
                                                                                                                                  
The result looks a bit puzzling at the first glance, since the velocity component $V_C$ of the
ionised gas in the top layer of the cloud  increases with the ionising flux, as
 shown in Figure \ref{figv}, so that we should expect the angle $\delta$ 
decreases with the ionising flux,  which will favour the formation of type C rim morphology.
 However, if we consider the effect of the enhanced
shock effect due to increased ionising flux, the compression to the core
in the front hemisphere will be stronger.
Therefore the core will exert a stronger attractive force on its peripheral gas particles which 
then results in
a higher $V_G$ so that angle $\delta$ of the velocity remains basically unchanged although
the magnitude of the total velocity may increase, which would decrease the time needed for the
gas particles in the ear-like structures to collapse onto the symmetrical axis.
 The same consideration can explain the insensitivity of the morphological evolution of the
cloud to the decreased radiation flux  and the time needed for the
gas particles to collapse onto the symmetrical axis is found to be elongated.

\subsection{Confrontation with  observations}
The simulations on a set of  molecular clouds reveal that the morphological evolution of a
molecular cloud under the effect of the UV radiation does not necessarily go through 
all of the three morphologies and then end up at cometary type C morphology
as described by the previous models. Depending on
its initial gravitational state, a cloud could evolve
to any one of the three type BRCs.
The diversity of the evolutionary sequence of BRCs revealed from our modelling provides
a good prospect for us to  explore the possible explanations for
the questions raised from  BRCs' observations and presented in the introduction of this paper.
                                                                                                                                    
The modelling results for the molecular cloud A, B, C  under the same initial temperature and
ionising flux tell us that the evolutionary paths of all
molecular clouds would  firstly
pass the type A  morphology. However some of  molecular clouds would only evolve to a
quasi-equilibrium type A  morphology plus a highly
condensed core at the head of its front hemisphere, if their initial self-gravitational 
energy is high enough. 
Therefore we shall statistically be able to observe many more type A
 BRCs than type B or C BRCs in a random observation.
                                                                                                                                    
Our modelling results also shows that the UV radiation
triggered collapse of a cloud onto the
centre core in its front hemisphere could occur with any one of
the three morphologies, so  it is not surprising at all that the UV radiation triggered star formation
occurring in type A BRCs could be observed. 
                                                                                                                                    
Finally, we come to understand the spacial sequence of type A-B-C BRCs with
their distance from the centre star in the Ori OB 1 association \citep{ogura}.
We might think this morphological distribution of BRCs as the result of 
decreasing illuminating flux with distance. However as shown in the argument in Section 4.6, 
the morphological evolution of the
cloud is insensitive to the change of the UV radiation flux in the range discussed.
Based on the results derived from the simulations, 
we can understand this observation from two aspects.  

First, we  know that where a new star is formed
inside a giant molecular cloud  as a result of
local molecular gas collapse onto a central point,
the density  of the surrounding
gas can be reasonably assumed to decrease with the distance from the central star. It is 
also known that giant molecular clouds  are  very clump \citep{tatematsu}. Considering the
above two points  we can assume that the
densities of  molecular clumps surrounding a new star  
decrease with their distance $D$ from the central star.
If we further assume that the clumps in a giant molecular cloud are of
similar radius $R$, then the initial gravitational potential energy of a
clump decrease with their distance from the centre of the star for the gravitational
potential energy can be written as $\Omega \sim M^2/R \sim R^5 n^2$, where $n$ is the initial
number density of the clumps.  

The molecular
clumps closest to the central star could possibly evolve to a quasi-equilibrium type A 
BRC because they have highest
initial self-gravitational potential energies, while for those clumps farther away, they 
evolve from type A to 
a quasi-equilibrium type B BRC because of their moderate initial self-gravitational potential energies.
For those farthest away from the central star,
they evolve to pass type A and B morphologies, then to a  
cometary type C morphology due to their lowest initial self-gravitational potential energies.

Secondly, we already showed in Section 4.4 that at the borders of HII region, type A BRCs are favoured.   

 Therefore the spatial sequence of type A - B - C BRCs with their distance to the
central star can be seen as a manifestation of the mass density distribution of the very clumpy
giant molecular clouds. 

\section{Conclusions}
A three dimensional and comprehensive RDI model has been developed with the SPH technique, which
self-consistently includes the hydrodynamical evolution, the self-gravity of the cloud,
the energy evolution, radiation transferring and a basic chemical network.
                                             T
The investigation on the morphological evolutions of BRCs are carried out by employing the newly developed  
model in order to
understand a series of puzzles coming from BRCs observations. It is found that the morphological evolution of 
a molecular cloud is much more sensitive to its initial self-gravitational state 
than to other physical conditions such as the initial temperature
between 20 and 100 K, or the ionising radiation flux within one order of magnitude higher or lower than
$J_0$(Lyman). Our modelling results have revealed that there are three
different evolutionary prospects when a molecular cloud is  under the
effects of ionisation radiations.

For a molecular cloud of an initial mass 35 M$_{\sun}$ and initial temperature 60 K under the effects of
UV and  FUV radiation fields specified above, the simulation results revealed:  a)
if its initial radius is 0.5 pc, it will evolve to a quasi-stable type A BRC;
b) when its initial radius increases to 1.3 pc, its evolution will firstly pass a
type A  morphology and continue its
journey to a quasi-stable type B rimmed morphology; c) when the initial radius further increases to
1.9 pc, its
evolutionary process follows a sequence of type A$\rightarrow$B$\rightarrow$C morphologies. d) Further 
increasing  the cloud's radius to  4.7 pc,  the  evolution of the cloud also follows the   
sequence of type A $\rightarrow$B$\rightarrow$C, but the finally formed type C BRC will be totally 
photoevaporated. 

Further simulation results also show that the initial central condensation of the mass in a cloud
could change its morphological evolution in term of its initially uniform correspondence. At the
 border of HII regions, the formation of type A BRCs are favoured.     
When a  molecular cloud is set to a zero
initial self-gravitational potential energy
or with a low initial density, its morphological evolution sequence
is consistent with that by Lefloch's model which is a 2-D model and didn't
include the self-gravity of the cloud. 

With the knowledge obtained from the modelling, we are now able to confront with the questions raised 
by observations. Since all of the clouds evolve to pass a type A morphology and some
of the clouds with enough high initial gravitational potential energy would just evolve to a quasi-stable
type A  morphology, statistically many more type A BRCs
should be observed over a random observation. For those BRCs who evolve to a quasi-stable
type A rimmed morphology, if the core mass contained in their front hemisphere is high enough
for it to collapse, new star could be triggered to form. Star forming signal should be observed.
Furthermore  if the initial densities of the
molecular clumps in  Ori OB 1 associations decrease
with their distances from the centre of the ionising star, a spacial distribution of type A-B-C
 BRCs with distance from the central star could exist.

It is obvious that the inclusion of the self-gravitation in our model revealed
a diversity of the evolutionary sequence of a molecular cloud under the
effects of UV radiations.
The satisfactory explanations based on the modelling results to the observational puzzles state that  the
self-gravity of a molecular cloud indeed plays an important role in controlling the evolutionary
destiny of a molecular cloud and should not be neglected especially when we try to achieve a
whole picture of the dynamical evolutions of molecular clouds. 
This is because  the radiation induced shock dramatically
enhances the effect of the self-gravity of the clouds. Consequently, some modifications on the 
region classification in Lefloch's parameter space have been confidently derived. The boundary
between the D-type ionisation front and 'ionising flush' is extended from $\delta = \sqrt{10}$ to 23
as a result of the inclusion of the self-gravity of the molecular cloud.

\clearpage
\begin{figure}
\begin{center}
\resizebox{5cm}{6cm}{\includegraphics[180,120][600,550]{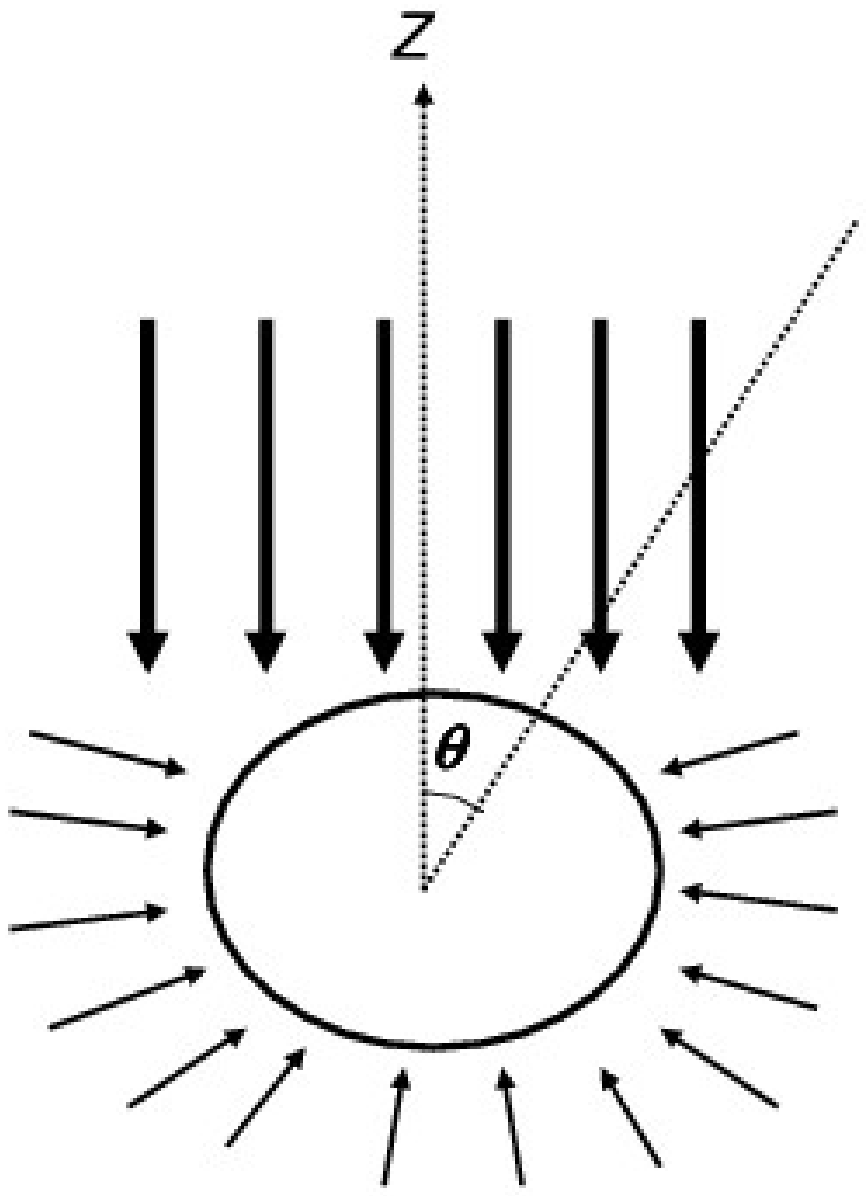}}
\caption{The initial geometry of a molecular cloud and the configuration of the radiation fields.
The heavy arrow lines above the upper hemisphere
represent the sum of the strong UV (Lyman continuum) flux from the massive stars above the molecular
cloud and the isotropic FUV flux from the background star radiation and the light arrow lines below the lower 
hemisphere represent
 the isotropic FUV radiation ($ h\nu < 13.6$ \, eV) flux only.}
\label{geometric}
\end{center}
\end{figure}

\clearpage
\begin{figure*}
\resizebox{16cm}{13cm}{\includegraphics{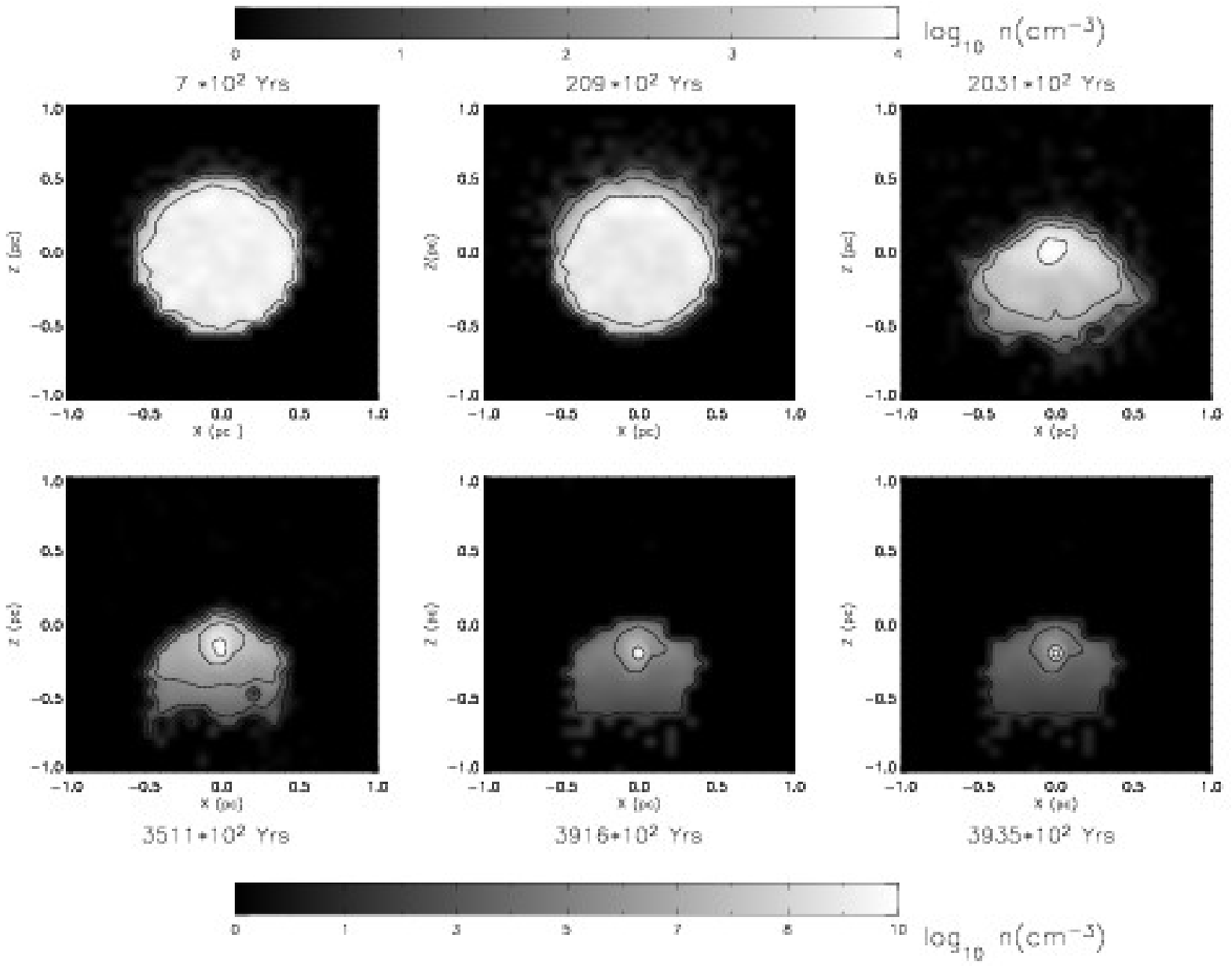}}
\caption{The evolutionary snapshots of the number density of the simulated molecular cloud A at
the cross section $y=0$
from  $t=700$ years to $t=0.394$ My. The top
grey bar shows the density scale in logarithm for the upper row snapshots and the bottom bar for the
second row snapshots.}
\label{figa1}
\end{figure*}

\clearpage
\begin{figure*}
\resizebox{16cm}{14cm}{\includegraphics{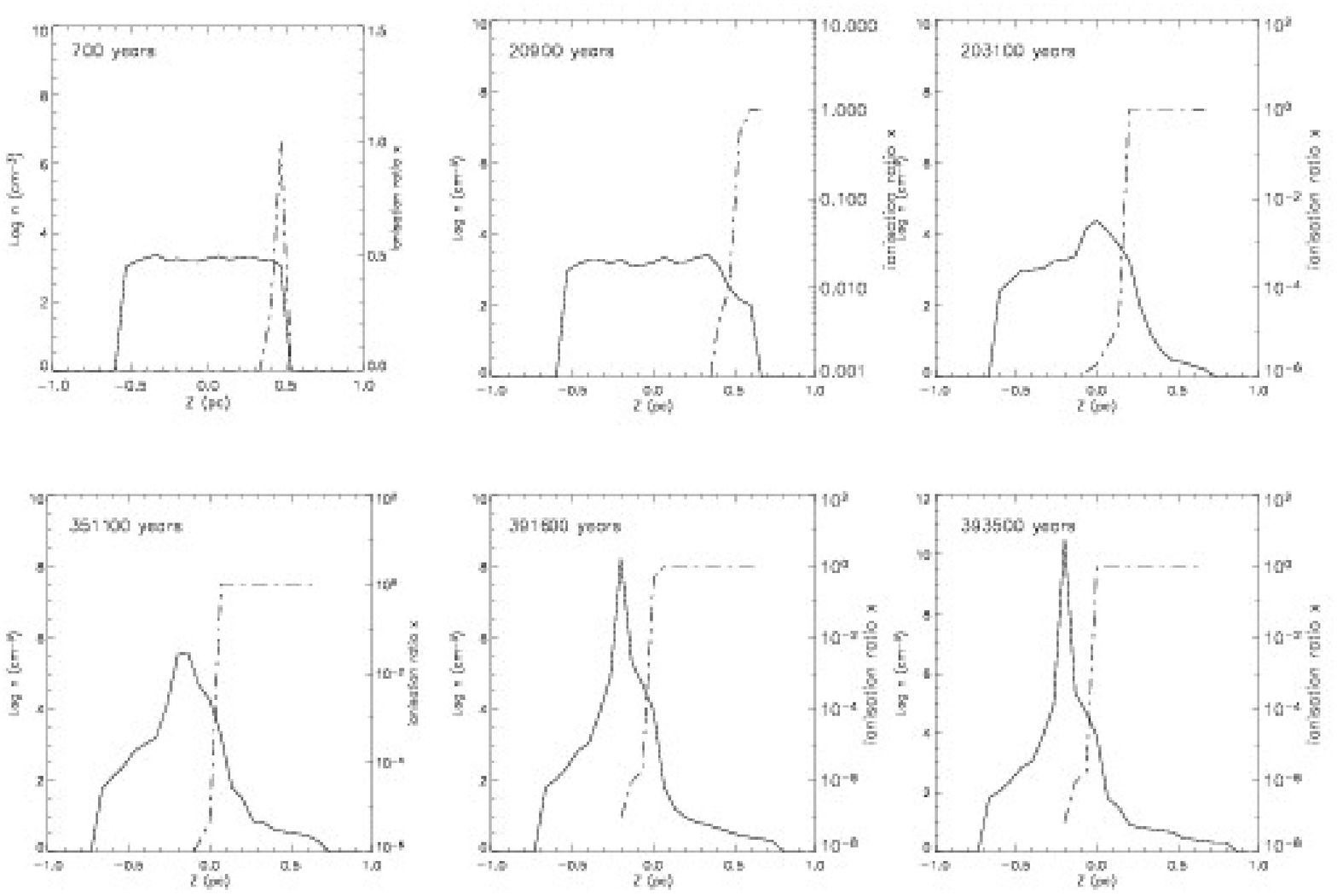}}
\caption{The distributions of the number density (solid line) and ionisation ratio x (dot-dashe line) 
of the 
simulated molecular cloud A along the symmetrical axis $z$ 
from  $t=700$ years to $t=0.394$ My.}
\label{figb1}
\end{figure*}

\clearpage
\begin{figure*}
\resizebox{16cm}{10cm}{\includegraphics{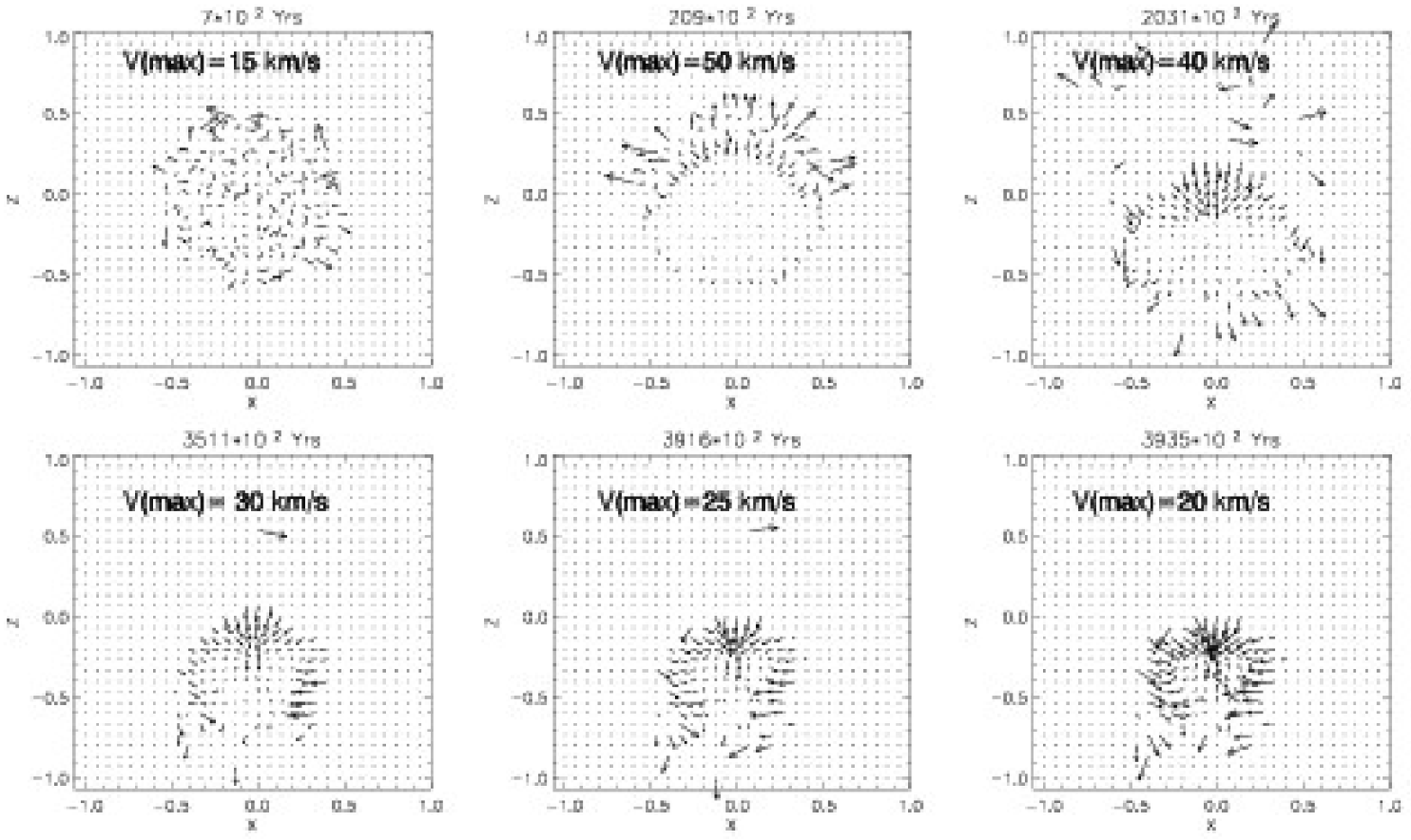}}
\caption{The evolution of the velocity  field for the same cloud as in Figure \ref{figa1} over a period of
 0.394 My.
The length of the arrow indicates the magnitude of velocity and the maximum magnitude of the
velocities at different stages of the evolution have been written inside the corresponding panels.}
\label{figc1}
\end{figure*}

\clearpage
\begin{figure*}
\begin{center}
\resizebox{18cm}{15cm}{\includegraphics{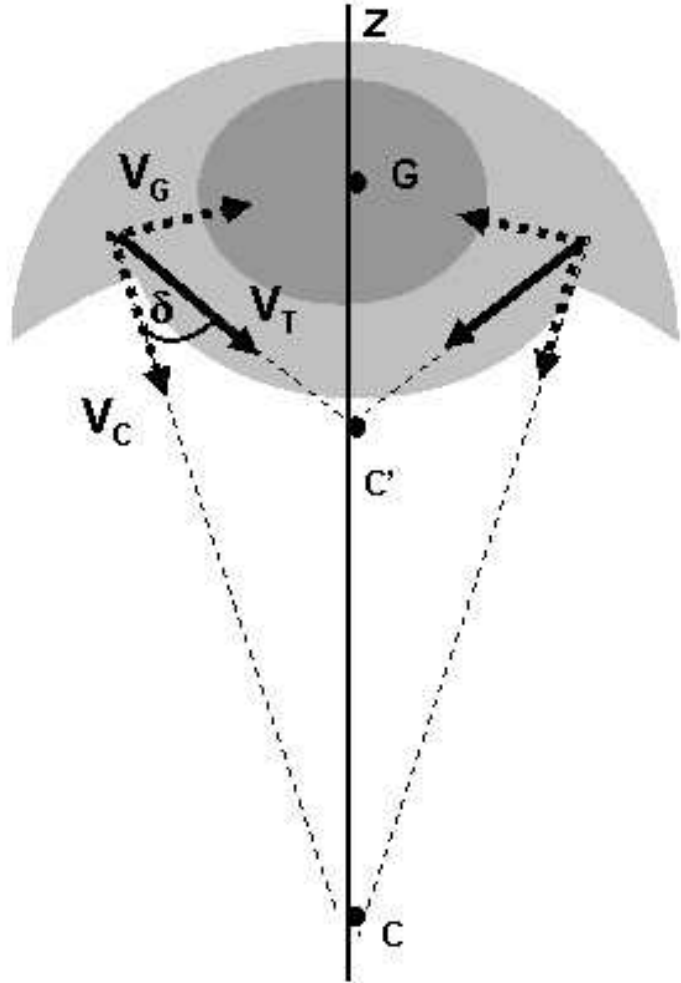}}
\caption{An illustration about the effect of the gravity of the main body of the cloud on
the velocity of the shocked gas particle in the 'ear-like' structures at the two sides of the
cloud. The central solid lines
are their structure symmetrical line $z$.}
\label{figv}
\end{center}
\end{figure*}

\clearpage
\begin{figure*}
\resizebox{16cm}{13cm}{\includegraphics{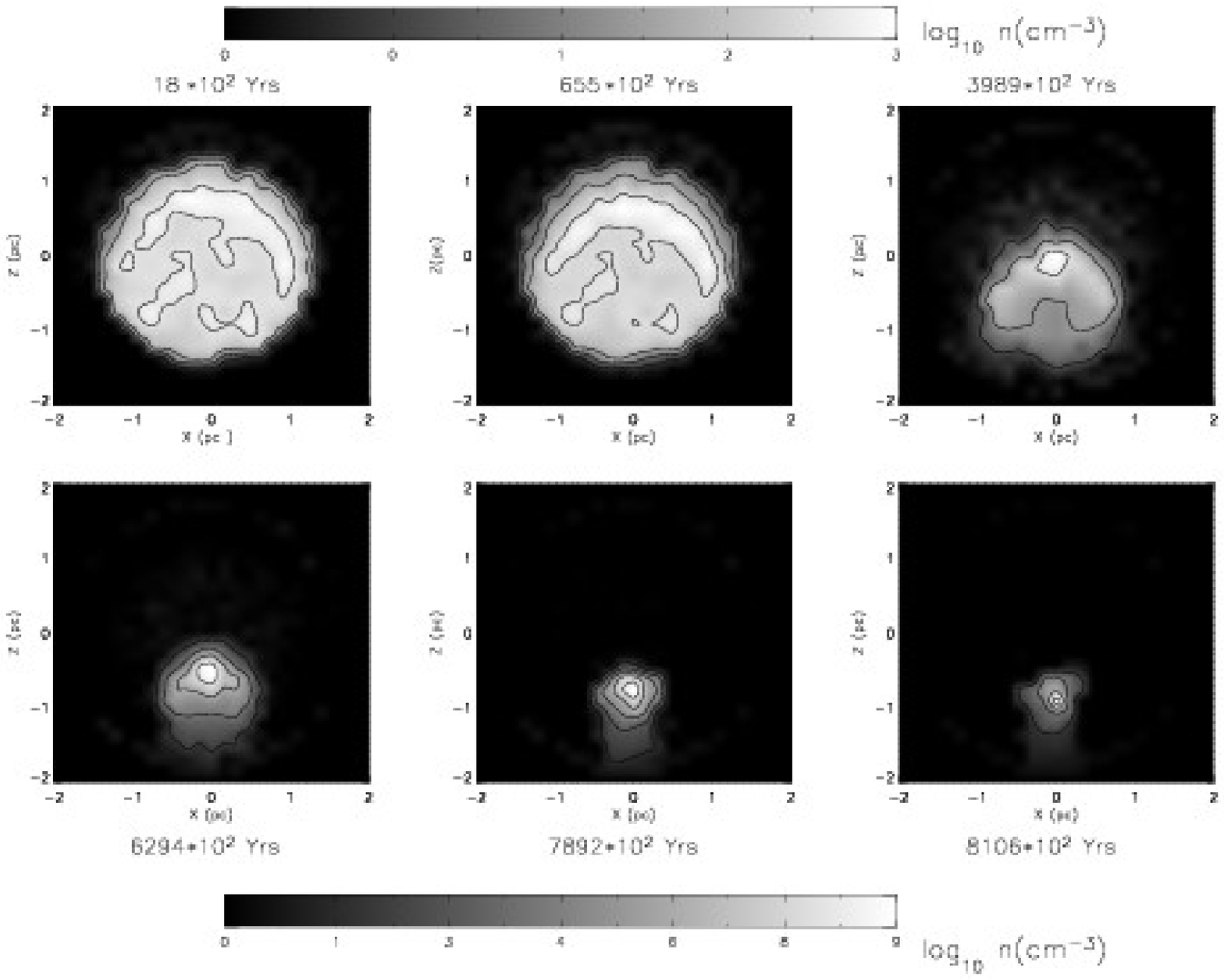}}
\caption{The evolutionary snapshots of the number density of the simulated molecular cloud B at
the cross section $y=0$
from  $t=1800$ years to $t=0.81 $ My. The top
grey bar shows the density scale in logarithm for the upper row snapshots and the bottom bar for the
second row snapshots.}
\label{figa2}
\end{figure*}

\clearpage
\begin{figure*}
\resizebox{16cm}{14cm}{\includegraphics{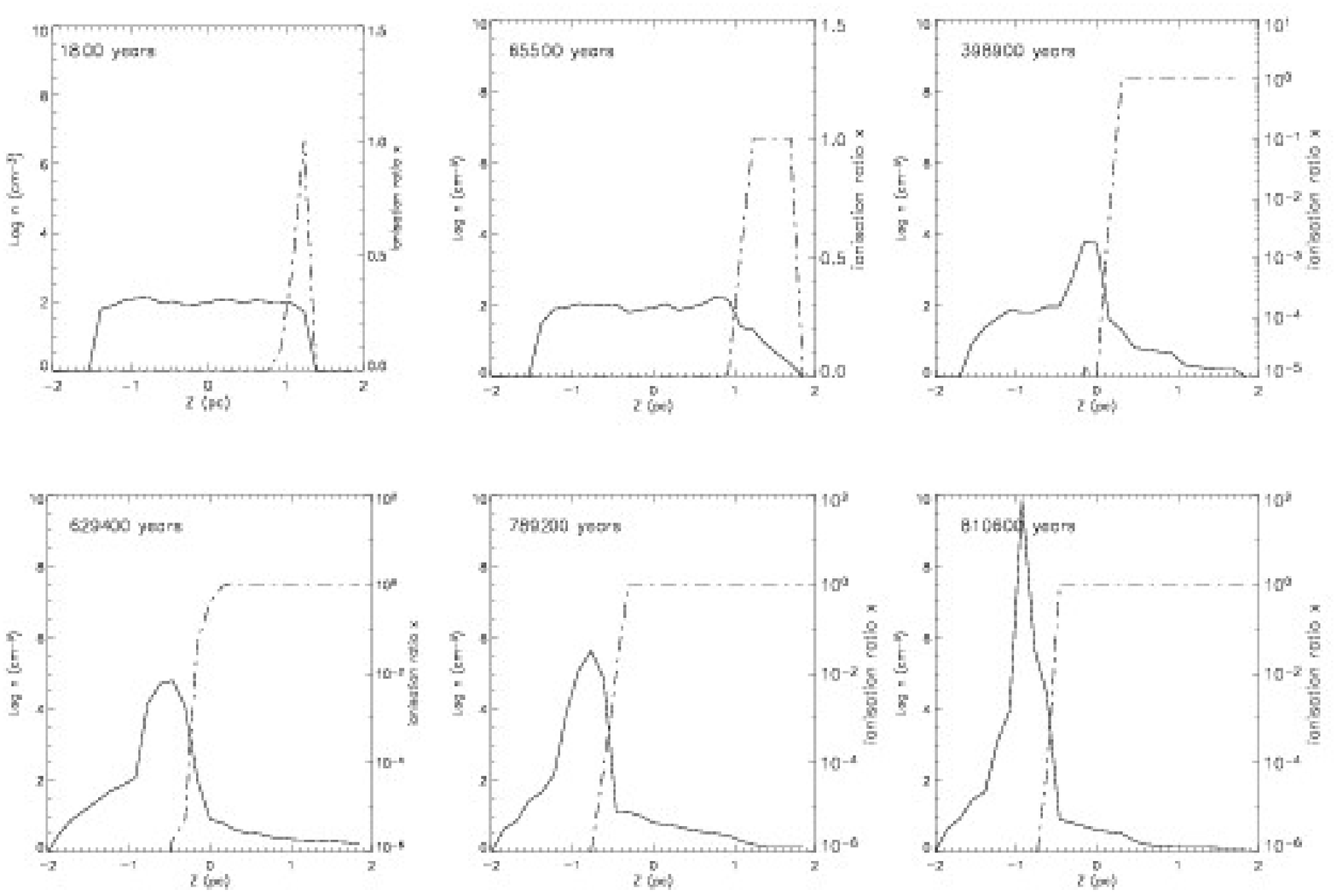}}
\caption{The distributions of the number density (solid line) and ionisation ratio x (dot-dashe line) 
of the simulated molecular cloud B along the symmetrical axis $z$ 
from  $t = 1800$ years to $t=0.81$ My.}
\label{figb2}
\end{figure*}

\clearpage
\begin{figure*}
\resizebox{16cm}{10cm}{\includegraphics{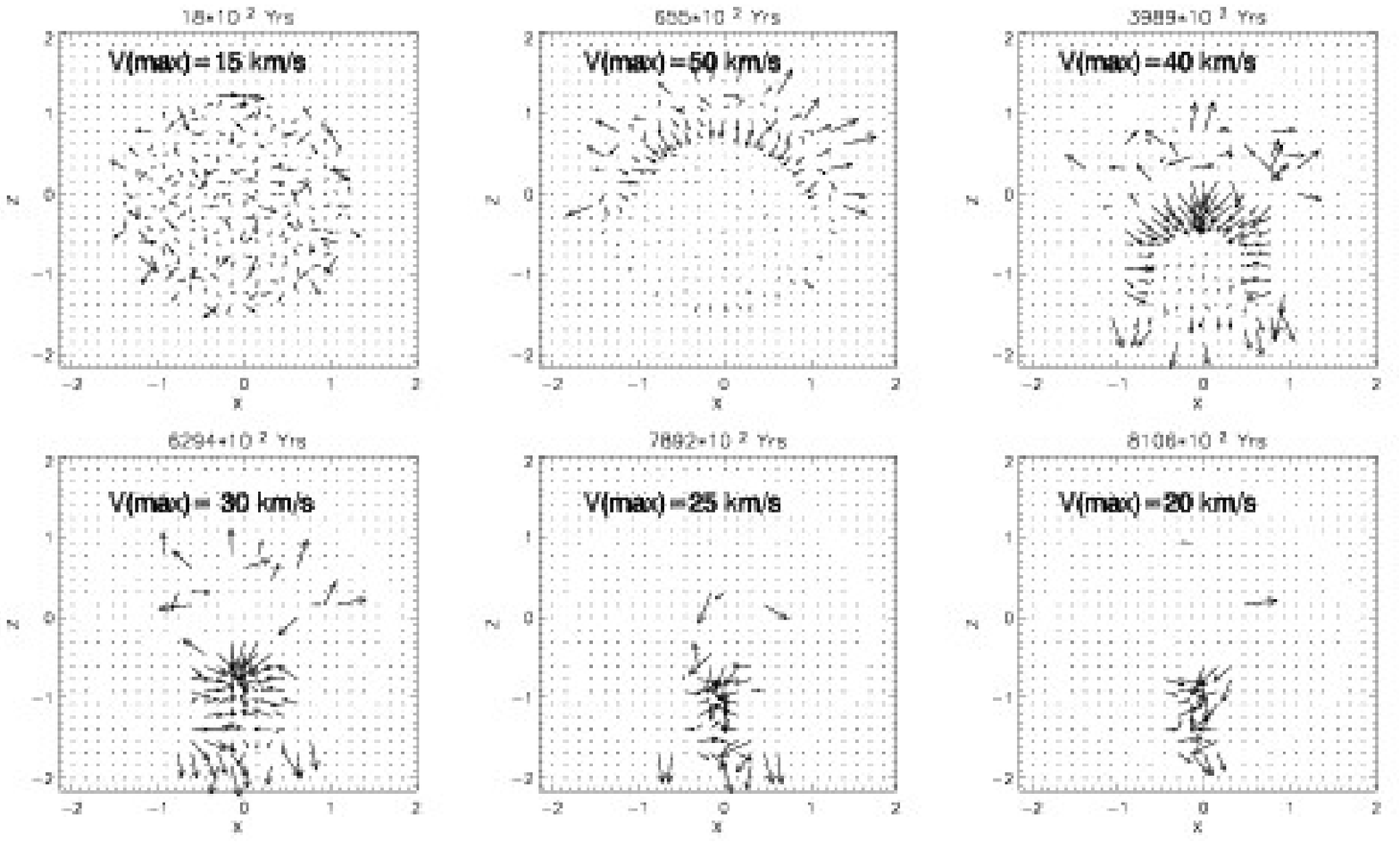}}
\caption{The evolution of the velocity  field for the same cloud as in Figure \ref{figa2} over a period of
time of 0.81 My. The length of the arrow indicates the magnitude of velocity and the maximum magnitude of the
velocities at different stages of the evolution have been written inside the corresponding panels.}
\label{figc2}
\end{figure*}

\clearpage
\begin{figure*}
\resizebox{16cm}{13cm}{\includegraphics{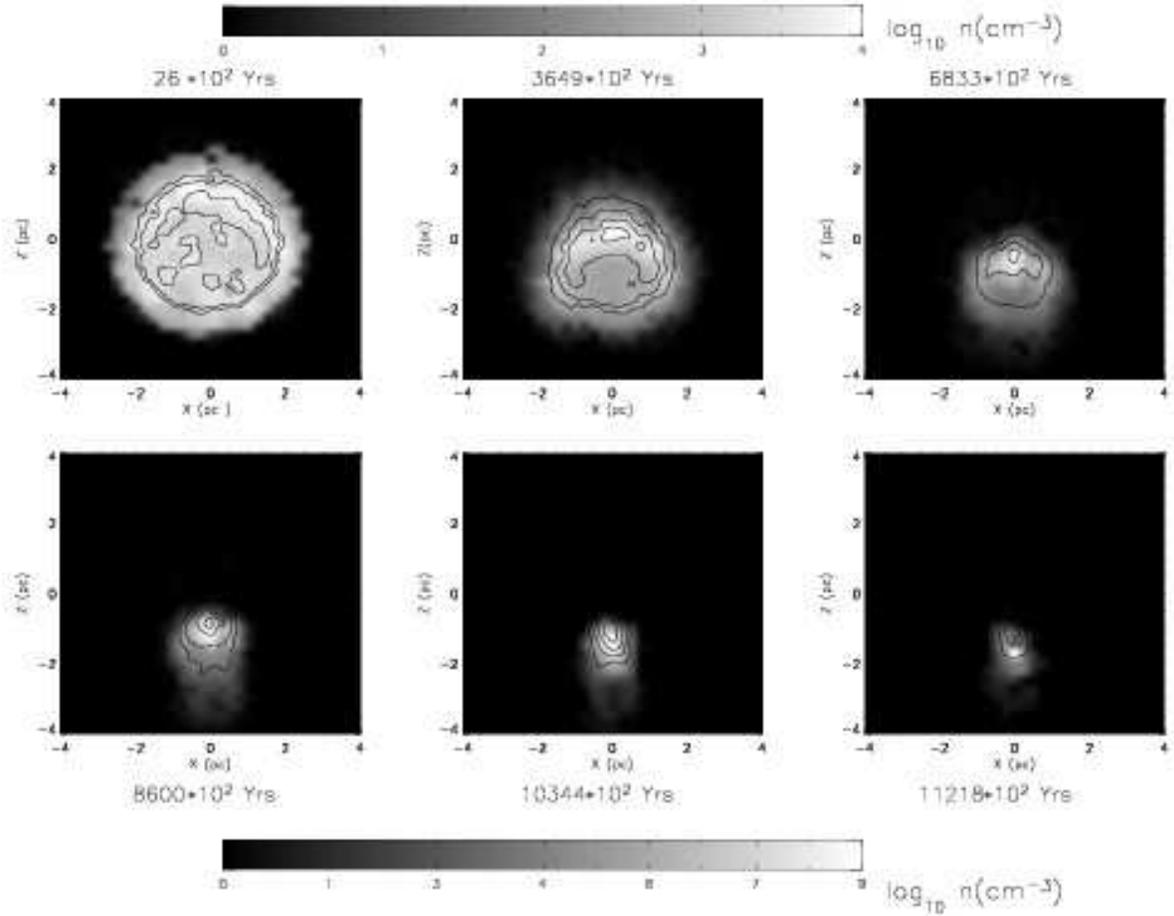}}
\caption{The evolutionary snapshots of the number density of the simulated molecular cloud C at
the cross section $y=0$
over a period of $1.12 $ My. The top
grey bar shows the density scale in logarithm for the upper row snapshots and the bottom bar for the
snapshots in the bottom row.}
\label{figa3}
\end{figure*}

\clearpage
\begin{figure*}
\resizebox{16cm}{14cm}{\includegraphics{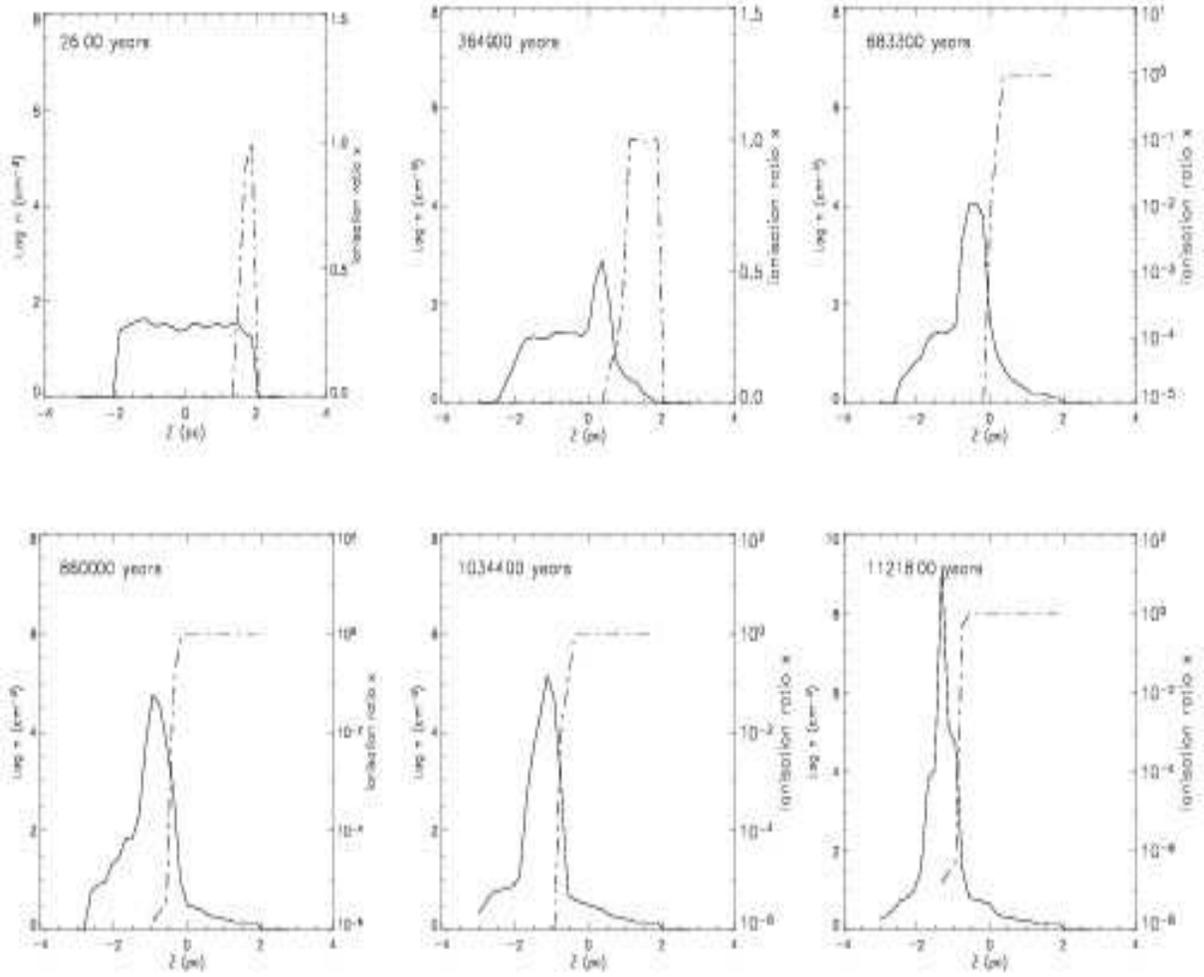}}
\caption{The distributions of the number density (solid line) and ionisation ratio x (dot-dashe line) 
of the 
simulated molecular cloud C along the symmetrical axis $x = 0$ 
over a period of 1.12 My.}
\label{figb3}
\end{figure*}

\clearpage
\begin{figure*}
\resizebox{14cm}{10cm}{\includegraphics{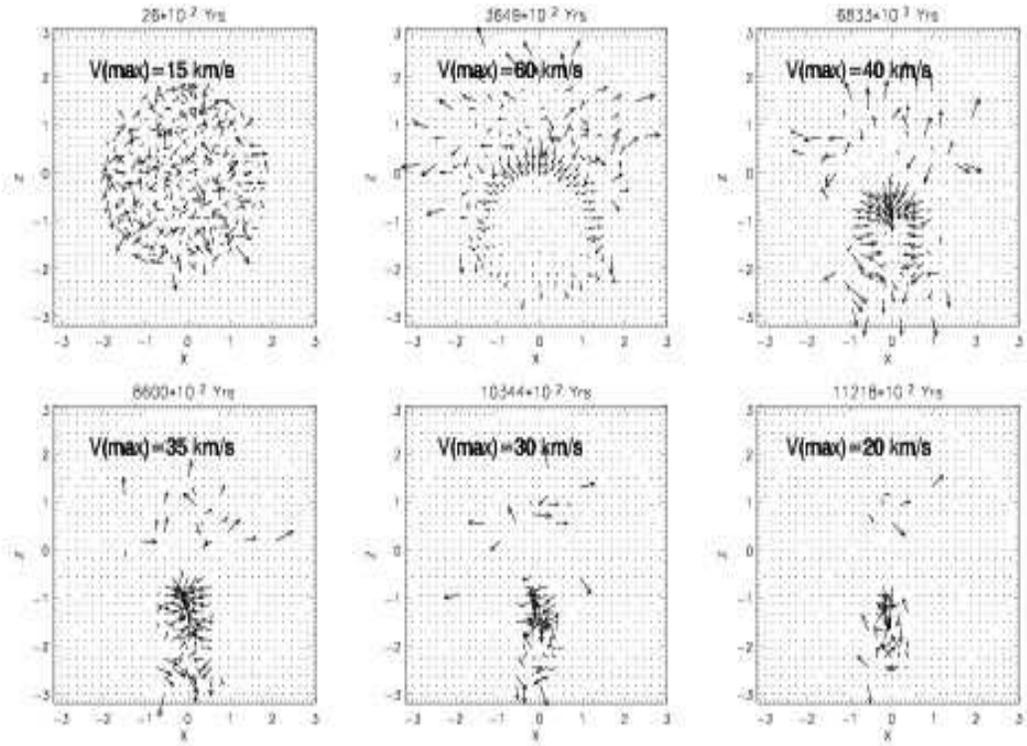}}
\caption{The evolution of the velocity field for the same cloud as in Figure \ref{figa3} over the same 
period of time. The length of the arrows indicates the magnitude of velocity and the longest
arrow in each panel represents
the maximum velocity in the corresponding velocity field distribution
at different stages of the evolution.}
\label{figc3}
\end{figure*}

\clearpage
\begin{figure*}
\resizebox{16cm}{15cm}{\includegraphics{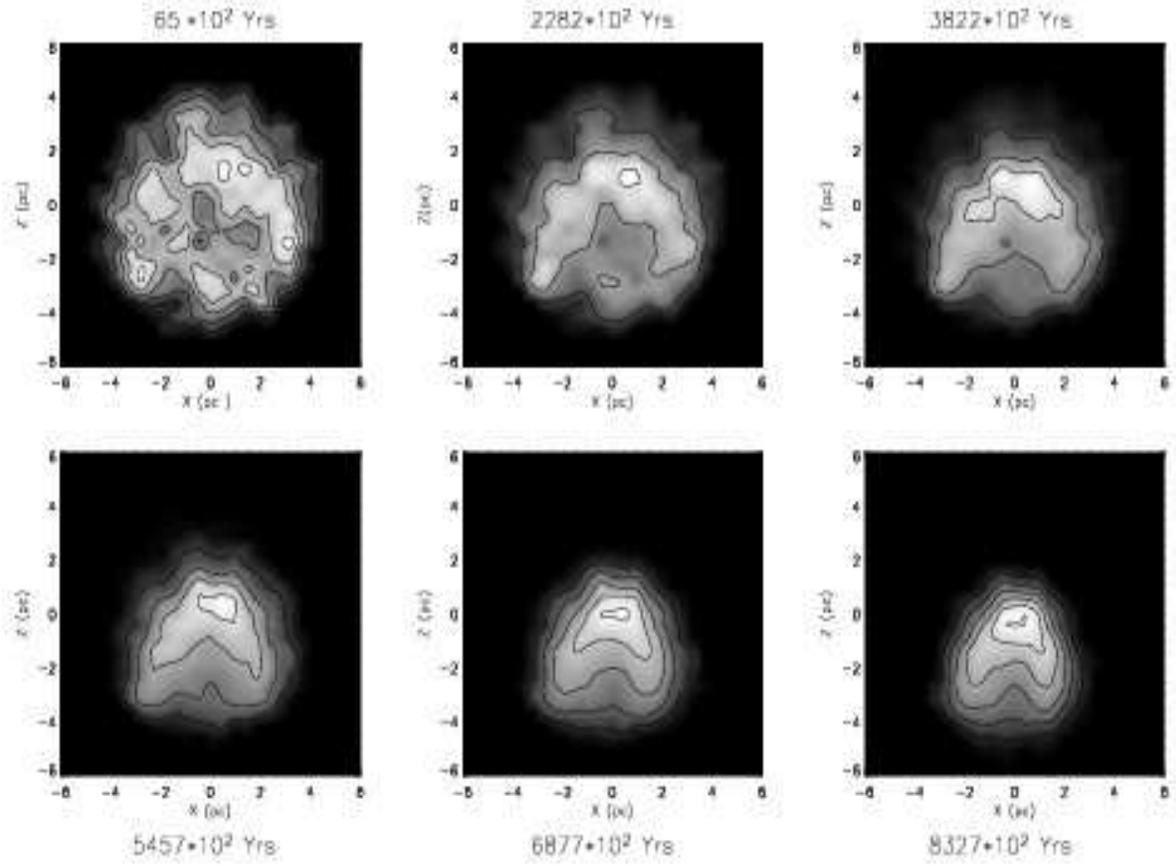}}
\caption{The growth and smooth processes of the large-scale surface instability 
over the evolutionary process in cloud D.}
\label{instability}
\end{figure*}

\clearpage
\begin{figure*}
\resizebox{14cm}{12cm}{\includegraphics{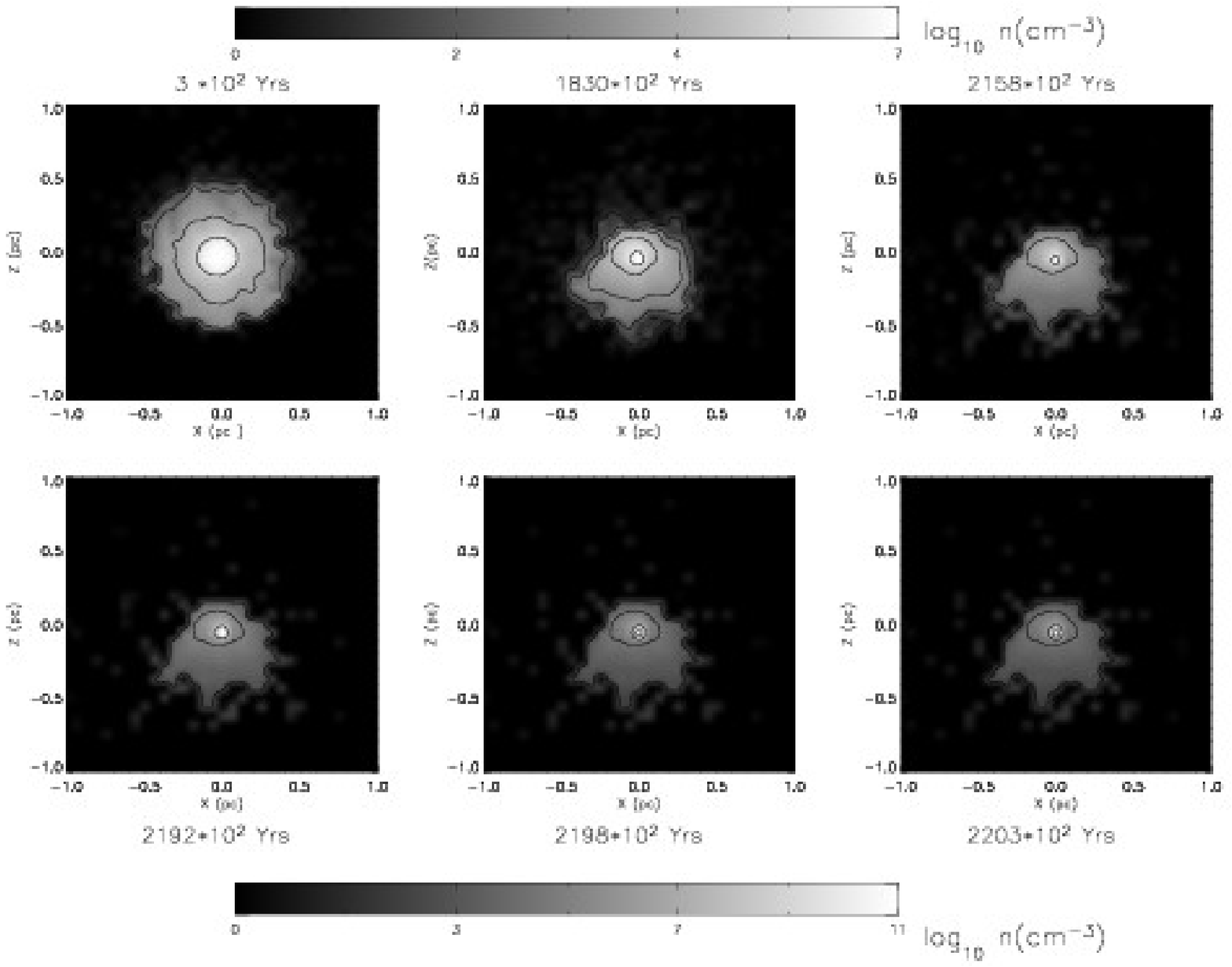}}
\caption{The evolutionary snapshots of the number density of the simulated molecular cloud A with 
an initial core-halo distribution  at
the cross section $y=0$.  The top
grey bar shows the density scale in logarithm for the upper row snapshots and the bottom bar for the
second row snapshots.}
\label{figa4}
\end{figure*}

\clearpage
\begin{figure*}
\resizebox{14cm}{12cm}{\includegraphics{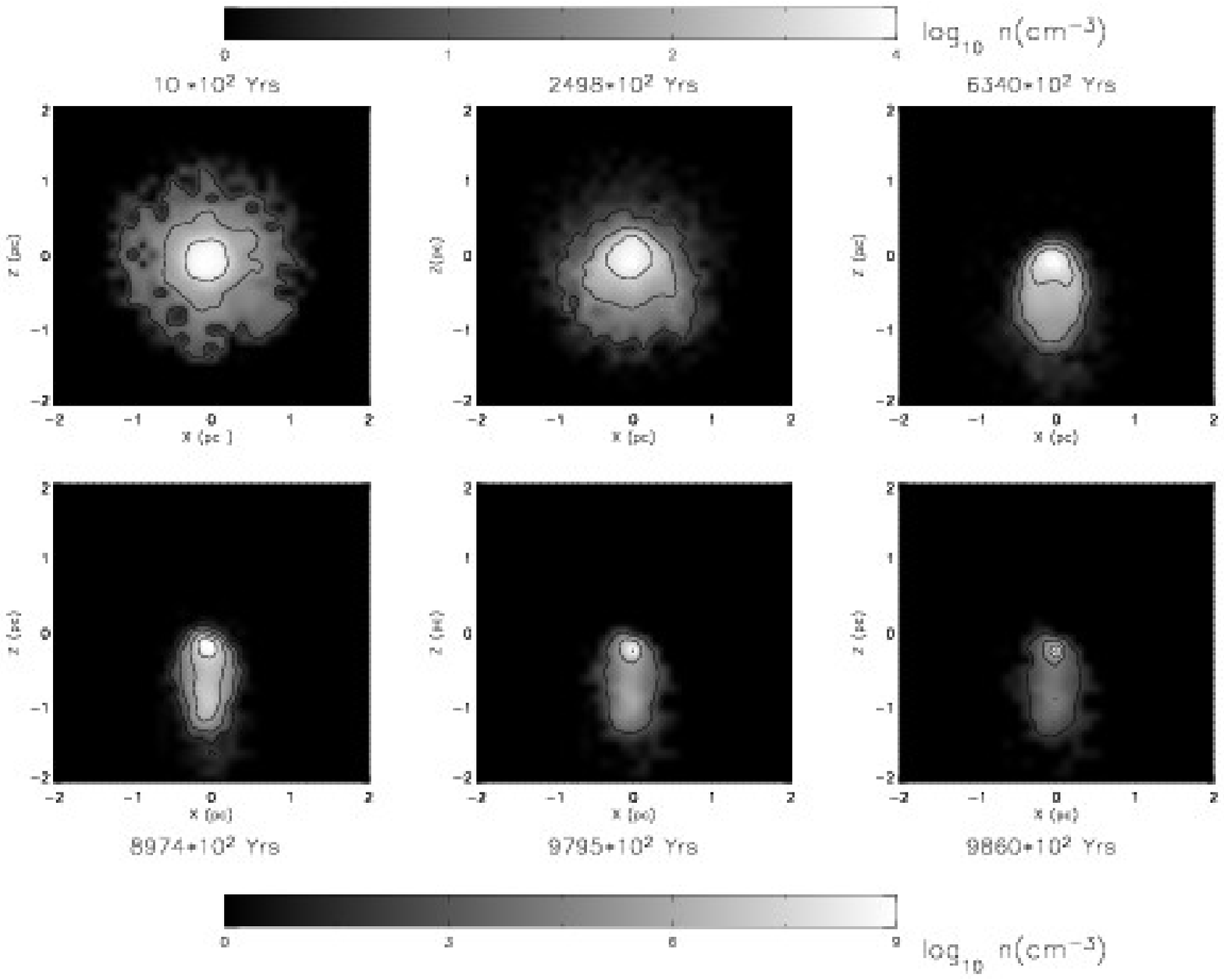}}
\caption{The evolutionary snapshots of the number density of the simulated molecular cloud B with
an initial core-halo distributionat
the cross section $y=0$. The top
grey bar shows the density scale in logarithm for the upper row snapshots and the bottom bar for the
second row snapshots.}
\label{figa5}
\end{figure*}

\clearpage
\begin{figure*}
\resizebox{14cm}{12cm}{\includegraphics{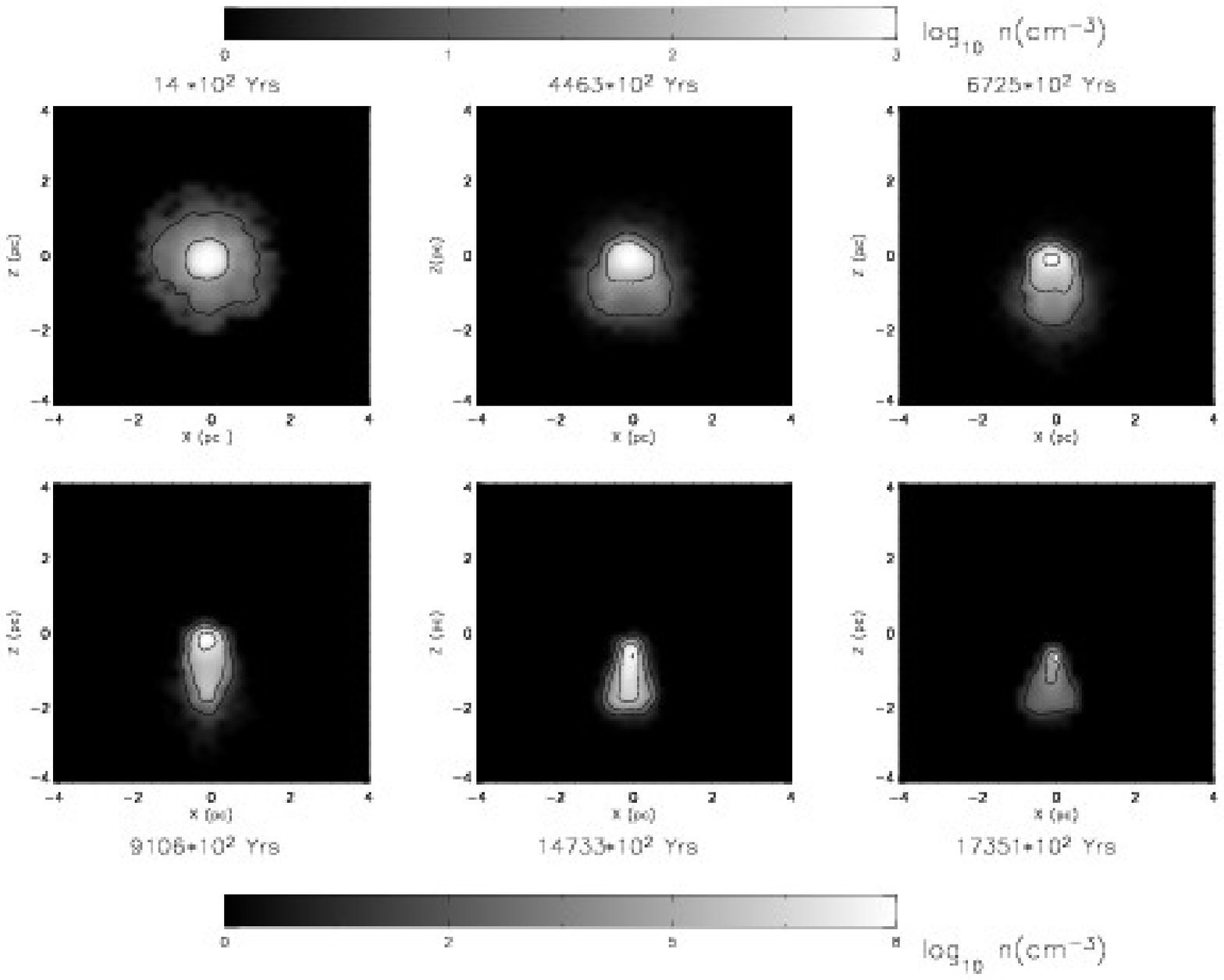}}
\caption{The evolutionary snapshots of the number density of the simulated molecular cloud C with
an initial core-halo distributionat
the cross section $y=0$. The top
grey bar shows the density scale in logarithm for the upper row snapshots and the bottom bar for the
second row snapshots.}
\label{figa6}
\end{figure*}

\clearpage
\begin{figure*}
\resizebox{16cm}{15cm}{\includegraphics{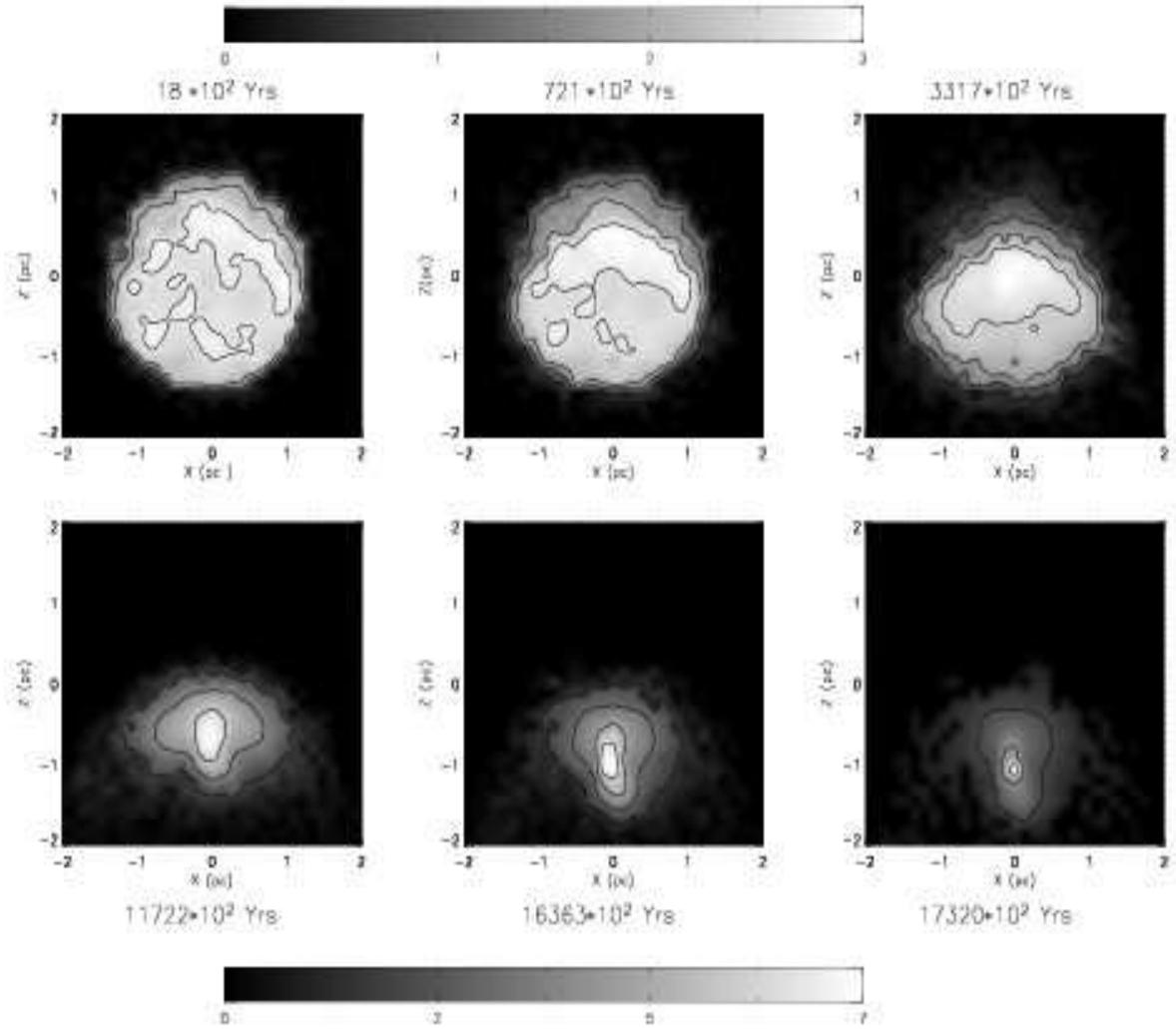}}
\caption{The evolutionary snapshots of the number density of the simulated molecular cloud B with a
zero external pressure boundary at the cross section $y=0$.  The top
grey bar shows the density scale in logarithm for the upper row snapshots and the bottom bar for the
snapshots in the bottom row.}
\label{figa7}
\end{figure*}

\clearpage
\begin{deluxetable}{lllllllll}
\tablecolumns{9}
\tablewidth{0pc}
\tablecaption{Parameters of set one clouds}
\tablehead{
\colhead{Cloud} & \colhead{$\Delta$} &  \colhead{$\Gamma$} & $n_0(cm^{-3})$ & 
\colhead{R(pc)} & \colhead{Region} & \colhead{M type\footnotemark[1]}
&\colhead{CFT\footnotemark[2]}&\colhead{CCT\footnotemark[3]}} 
\startdata
A & 0.68 & 55 & 2672 & 0.5 & V & A & 0.35 & 0.39 \\ 
B & 7.3  & 89.1 & 152 & 1.3 & III & B & 0.63 & 0.6\\
C & 18 & 108 & 49 & 1.9 & III & C & 0.86 & 1.12 \\
D &182 & 169 & 3.2 & 4.7 & III & C & 1.0 & N/A     
\enddata
\end{deluxetable} 
\footnotetext[1]{Morphology type}
\footnotetext[2]{Core formation time in My}
\footnotetext[3]{Core collapse time in My}

\clearpage
\begin{deluxetable}{lllllllll}
\tablecolumns{9}
\tablewidth{0pc}
\tablecaption{Parameters of second set clouds}
\tablehead{
\colhead{Cloud} & \colhead{$\Delta$} & \colhead{$\Gamma$} & $n_0(cm^{-3})$ & 
\colhead{M(M$_{\sun}$}) & \colhead{Region} & \colhead{M type\footnotemark[1]}
&\colhead{CFT\footnotemark[2]}&\colhead{CCT\footnotemark[3]}} 
\startdata
A & 0.68 & 55 & 2672 & 35 & V & A & 0.35 & 0.39 \\ 
A' & 1.94 & 55 & 916 & 12 & V & B & 0.45 & 0.55  \\
A'' & 2.9 & 55 & 610 & 8 & III & C & 0.48 & 0.72 \\
A''' & 23 & 55 & 76 & 1 & III & C & 0.25 & N/A
\enddata
\end{deluxetable} 
\footnotetext[123]{They have the same meanings as that in Table 1.}

\end{document}